\documentclass[fp,twocolumn]{jpsj3}
\usepackage{txfonts}

\usepackage[dvipdfmx]{graphicx}
\usepackage{color}
\usepackage{amssymb}
\usepackage{amsmath} 
\usepackage{ascmac}

\usepackage{braket}
\usepackage{mathrsfs}

\usepackage{floatflt}
\usepackage{nidanfloat}

\newcommand{\tabref}[1]{Table~\ref{#1}}
  \newcommand{\equref}[1]{Eq.~(\ref{#1})}
    \newcommand{\equsref}[1]{Eqs.~(\ref{#1})}
  \newcommand{\figref}[1]{Fig.~\ref{#1}}
    \newcommand{\figsref}[1]{Figs.~\ref{#1}}

\newcommand{\scr}[1]{\mathscr{#1}}
\newcommand{\bi}[1]{\ensuremath{\boldsymbol{#1}}}

\newcommand{\D}{\Delta}

\newcommand{\la}{\lambda}
\newcommand{\ve}{\varepsilon}
\newcommand{\om}{\omega}
\newcommand{\Om}{\Omega}
\newcommand{\al}{\alpha}

\newcommand{\ga}{\gamma}
\newcommand{\Ga}{\Gamma}
\newcommand{\si}{\sigma}

\newcommand{\ka}{\kappa}
\newcommand{\lavf}{\lambda_{\varepsilon_{\rm F} }}

\title{Relativistic correction to the magnetoresistance of the Dirac electron system}

\author{Mitsuaki Owada and Yuki Fuseya}
\inst{Department of Engineering Science, University of Electro-communications } 

\abst{%
The magnetoresistance (MR) for relativistic electrons, i.e., the Dirac electrons in solids, is investigated on the basis of the Boltzmann's theory. The new formula of MR so obtaiend includes a relativistic correction, which has not appeared in the conventional Boltzmann's theories. Our formula is quantitatively consistent with that obtained by the Kubo theory except for the quantum oscillations. While a field dependence of MR is unclear in the formula by Kubo theory, a clear field dependence is obtained in our formula, which is useful for the analysis of experimental results. The effects of the relativistic correction on the MR for the one-band and two-band model are discussed. By taking into account the field dependence of carrier density in semimetals, the linear field dependence of MR is explained by our formula based on the Boltzmann's theory.}

\begin{document}
\maketitle

\section{Introduction}
\begin{table*}[b]
  \centering
    \caption{Summary of the relativistic effects on the MR ($\rho_{xx}$), the magneto Hall resistivity ($\rho_{yx}$), and the Hall coefficients ($R_{\rm H}$). $\tilde{\rho}_{ij}$ and $\tilde{R}_{\rm H}$ are relativistic quantities renormalized with the conventional one. Here, $c = n/p$ ($n$ is the carrier density of electrons and $p$ is that of holes), $\ka = \mu_0 / \nu_0$ ($\mu_0$ and $\nu_0$ are the mobility of electrons and holes, respectively). $A_1 = (1 + c \ka )^2/(1-c \ka^2)$ and $A_2  = (1+\ka)/(1-\ka)$.  \label{tab_la_B_res}}

\begin{tabular}{|c|c|c|c|c|c|c|c|c|cc|} \hline
& \multicolumn{3}{|c|}{One band model} & \multicolumn{6}{|c}{Two band model}  & \hspace{-5mm}  \\ \cline{2-10} 
& &  &  & \multicolumn{3}{|c|}{$ n \neq p$}  & \multicolumn{3}{|c}{$n = p$} & \hspace{-5mm}  \\ \cline{5-10}
& $\tilde{\rho}_{xx}^{\rm R}$ & $\tilde{\rho}_{yx}^{\rm R}$  & $\tilde{R}_{\rm H}^{\rm R}$ & & & & & & & \vspace{-5pt}\hspace{-5mm}  \\
& & &  &  $\tilde{\rho}_{xx}^{\rm R+NR}$ & $\tilde{\rho}_{yx}^{\rm R+NR}$  & $\tilde{R}_{\rm H}^{\rm R+NR}$ &  $\tilde{\rho}_{xx}^{\rm R+NR}$ & $\tilde{\rho}_{yx}^{\rm R+NR}$  & $\tilde{R}_{\rm H}^{\rm R+NR}$ & \vspace{-5pt} \hspace{-5mm}  \\ 
&&&  & & &   & &  & & \vspace{-1pt} \hspace{-5mm}  \\
 &&&  & &\multicolumn{2}{c|}{ }   & &  \multicolumn{2}{c}{} & \vspace{-9pt} \hspace{-5mm}  \\ \hline
  &&&  & & \multicolumn{2}{c|}{ }  & &  \multicolumn{2}{c}{} & \vspace{-5pt} \hspace{-5mm}  \\
Weak fields & & & & &  \multicolumn{2}{c|}{ }  & &  \multicolumn{2}{c}{} & \hspace{-5mm}  \\
  &&&  & & \multicolumn{2}{c|}{ }  & &  \multicolumn{2}{c}{} & \vspace{-9pt} \hspace{-5mm}  \\
($\la_{\ve_{\rm F}}^2 \mu_0^2B^2 \ll 1, \nu_0^2B^2  \ll 1$)   & & & & & \multicolumn{2}{c|}{ } & &  \multicolumn{2}{c}{} & \vspace{-22pt}\hspace{-5mm}  \\
&&&& $\displaystyle \frac{ 1 + c \ka }{ 1 + c \ka \lavf }$ &  \multicolumn{2}{c}{ $ \displaystyle A_1 \times  \frac{ 1 - c \ka^2 \lavf^2 }{ (1 + c \ka \lavf )^2  } $ }  & $\displaystyle \frac{ 1 + \ka }{ 1 + \ka \la_{\ve_{\rm F}} }$ &  \multicolumn{2}{c}{}  &  \vspace{-5pt}\hspace{-5mm}  \\  
  &&&  & & \multicolumn{2}{c|}{ } & &  \multicolumn{2}{c}{} & \hspace{-5mm}  \\  \cline{1-1} \cline{5-8} 
  &&&  & & \multicolumn{2}{c|}{ } & &  \multicolumn{2}{c}{} & \vspace{-20pt} \hspace{-5mm}  \\
 &$\displaystyle \frac{1}{\la_{\ve_{\rm F}}}$& $1$ & $1$  & &  \multicolumn{2}{c}{ }  & & \multicolumn{2}{c}{ $\displaystyle  A_2 \times \frac{ 1 - \ka \la_{\ve_{\rm F}}  }{ 1 + \ka \la_{\ve_{\rm F}}  }  $   }  &  \hspace{-5mm}  \\
 &&&  & & \multicolumn{2}{c}{ } & &  \multicolumn{2}{c}{} & \vspace{-16pt} \hspace{-5mm}  \\ 
Strong fields   &&&  & & \multicolumn{2}{c|}{ } & &  \multicolumn{2}{c}{} & \vspace{-7pt} \hspace{-5mm}  \\
 &  && & $\displaystyle  \frac{ c + \ka \lavf }{ ( c + \ka )\lavf }  $ & \multicolumn{2}{c|}{ 1 }  & $\displaystyle \frac{ ( 1 + \ka )  \lavf}{ 1 + \ka \lavf}$ &   \multicolumn{2}{c}{ } & \vspace{-12pt} \hspace{-5mm}  \\
 ($\la_{\ve_{\rm F}}^2 \mu_0^2B^2 \gg 1,  \nu_0^2B^2  \gg 1$ )& & & & & \multicolumn{2}{c|}{ } & &  \multicolumn{2}{c}{} & \vspace{-5pt} \hspace{-5mm}  \\ 
& & & & & \multicolumn{2}{c|}{ } & &  \multicolumn{2}{c}{ }  & \hspace{-5mm}  \\ \hline
    \end{tabular}
\end{table*}

Linear magnetoresistance was first reported by Kapitza in 1928 \cite{Kapitza1928}. In the following year, he showed the magnetoresistance (MR) in various materials\cite{Kapitza1929p1,Kapitza1929p2}. MR was actively researched after his studies\cite{Sommerfeld1931,Harding1933,Jones1934,Schubnikow1935,Milner1937,Davis1939}. Currently, MR is still utilized to understand the physical properties of solids. In contrast, an extremely large magnetoresistance was reported in the semimetal $\rm WTe_2$, recently\cite{Ali2014}. Following this report, the nature of large MR has been studied in many materials that have strong spin-orbit couplings. Especially, Dirac electron systems, in which the energy dispersion is linear in the wave number $k$, have been the focus of various studies \cite{Wang2014,Feng2015,Shekhar2015,Novak2015,He2016,Shrestha2017,Zhu2015,Zhu2017,Fauque2009NJP,Fauque2009PRB,Zhu2011NatPhys,Collaudin2015}.

The properties of MR is governed by the mobile electrons, so that it strongly depends on the energy dispersion of electrons. In most cases, the analysis of the MR has been based on the theory for the non-relativistic electrons, where the dispersion is quadratic in $k$. However, the MR of relativistic electrons, where the dispersion is linear (Dirac electrons), has not been examined in detail. Furthermore, in the case of semimetals, which exhibit large MR, relativistic and non-relativistic carriers can coexist, we need much more complex analysis. 
For example, it is well known that the MR increases with the square of the magnetic field ($\rho_{xx} \propto B^2$) \cite{Gropara}, when there are non-relativistic electron and hole carriers (NR+NR). Then, is this field dependence the same for relativistic carriers, such as (R+R) and (R+NR)? Actually, bismuth, a typical semimetal, can be classified into (R+NR) type\cite{Dresselhaus1971,Edelman1976,Issi1979,Fuseya2015a}, and the origin of its (quasi) linear MR is still controversial. Here, ``R'' and ``NR'' denote relativistic  and non-relativistic, respectively.

In previous studies, the contribution of various effects for Dirac electrons was examined, e.g., the Coulomb screening effect, impurity potential, and quantum limit, \cite{Peres2007,Sinitsyn2007,Wang2012_linear,Abrikosov1998}. However, these papers do not discuss the effects of the relativistic dispersion. In addition, these theories are applicable under some restricted conditions. It is difficult to discuss the MR of any magnetic field with these theories.

What makes the analysis of MR very complex is the fact that the experimentally obtained MR, $\hat{\rho}$, is given in the tensor form. Theoretically, on the other hand, the conductivity tensor, $\hat{\sigma}$, is first obtained, and then, we have to calculate the inverse tensor of $\hat{\sigma}$. Therefore, in order to analyze the experimental data of MR, it is of the prime importance to obtain a clear field dependence not of $\hat{\sigma}$ but of $\hat{\rho}$. For this purpose, the semiclassical approach based on the Boltzmann equation is very powerful. On the other hand, the approach based on the Kubo formula is difficult to see the explicit field dependence of $\hat{\rho}$, so that it is rather hard to analyze the experimental data, even though the result is rigorously quantum.

In this paper, we firstly obtain the formula of MR for the relativistic electrons with the Wolff model, which is the effective model of Dirac electron systems, such as bismuth\cite{Wolff1964, Fuseya2015a}, based on the Boltzmann equation. We found a relativistic correction factor, $\la_{\ve} = \D/\ve$, which does not appear in the conventional formula for non-relativistic carriers ($\D$ is a half of the band-gap, $\ve$ is the energy). 
The relativistic correction $\la_{\ve}$ makes change the amplitude of $\rho_{ij}$, e.g., $\rho_{xx}^{\rm R} = \lavf \rho_{xx}^{\rm NR}$ ($\rho_{xx}^{\rm R}$ and $\rho_{xx}^{\rm NR}$  are the relativistic and non-relativistic magnetoresistivity in the one band model, respectively). 
The $\la_{\ve_{\rm F}}$-dependence of $\rho_{ij}^{\rm R, R+NR}$ and the Hall coefficients $R_{\rm H}^{\rm R,R+NR}$ for one-band and two-band models are summarized in \tabref{tab_la_B_res}. For the quantitative analysis of experimental MR, this correction term plays a crucial role. It is also shown that the field dependence of $\hat{\rho}$ is very clear in our formula and it is essentially consistent with that based on the Kubo formula. In addition, we succeed in explaining the linear MR by taking into account the field dependence of carrier density based on the effective models.

The rest of this paper is organized as follows. Section 2 describes the formulation of the relativistic conductivity. In Sect. 3 to Sect. 5, we calculate the MR based on Boltzmann theory. The MR with the one-band and two-band model are described in Sects. 3 and 4, respectively, and the MR under the quantum limit is described in Sect. 5. In Sect. 6, the magnetoconductivities based on the Boltzmann are compared with that based on the Kubo theory. Our conclusion is presented in Sect. 7.

\section{Relativistic conductivity $\hat{\si}^{\rm R}$}
We derive the relativistic conductivity $\hat{\si} ^{\rm R}$ for the anisotropic Wolff model. The hamiltonian is given by\cite{Wolff1964,Fuseya2015a}:
\begin{align}
\label{eq_wolff}
\scr{H} 
= 
\D \ga_{4} + i\hbar \bi{k}\cdot\left[  \sum_{i = 1}^3 \bi{W}(i) \ga_4 \ga_{i} \right],
\end{align}
where $\bi{k}$ is the wave vector measured from an extremum of dispersion. $\ga_i$ is the $4\times 4$ Dirac matrix of the form
\begin{align}
\label{eq_dirac_mat}
  \ga_{i=1,2,3} &= 
\left(
\begin{array}{cc}
0 & \si_{i} \\
\si_{i} & 0 \\
\end{array}
\right), \\
 \ga_4 &=
\left(
\begin{array}{cc}
I & 0 \\
0 & -I \\
\end{array}
\right), 
\end{align}
where $\si_i$ is the Pauli spin matrix. $\bi{W}(i)$ is related to the matrix elements of the velocity operator for the same spin, $\bi{t}$, and for the opposite spin, $\bi{u}$, as
\begin{align}
\label{eq_W}
\bi{W}(1) &= {\rm Im}(\bi{u}), \\
\bi{W}(2) &= {\rm Re}(\bi{u}),\\
\bi{W}(3) &= {\rm Im}(\bi{t}).
\end{align}
The eigenvalue of \equref{eq_wolff} is 
\begin{align}
\label{energy_dirac0}
\pm \ve ^{\rm R}
= \pm \sqrt{\D^2 + \D \hbar^2 \bi{k}\cdot \hat{\al}\cdot  \bi{k}  },
\end{align}
where $\hat{\al}$ is the inverse mass tensor given by $\al_{ij}=\left( \sum_\mu W_i (\mu) W_j (\mu) \right)/\Delta$.
The velocity of $\ve^{\rm R}$, i.e., the relativistic velocity $\bi{v}^{\rm R}$ is given as
\begin{align}
\label{eq_v_rela}
\bi{v}^{\rm R}
  &= \frac{1}{\hbar}\frac{\partial \ve^{\rm R}}{\partial \bi{k}} 
= \la_{\ve} \bi{v}^{\rm NR},  \\
\la_{\ve}   &=  \frac{\D}{\ve^{\rm R}}.
\end{align}
Here, $\bi{v}^{\rm NR}$ is the velocity of the non-relativistic carriers defined by
\begin{align}
\bi{v}^{\rm NR}   &=   \hat{\al} \cdot \hbar \bi{k},
\end{align}
using the inverse mass tensor $\hat{\al}$. The dimensionless parameter $\la_{\ve_{\rm F}}$ expressed the relativistic correction.

The current density $\bi{j}$ is described as follows\cite{Wilson1965,Jones1934,Davis1939,Abeles1954}: 
\begin{align}
\label{eq_current}
\bi{j} &= \frac{e}{(2 \pi ) ^3} \int \bi{v} \Phi \frac{\partial f_0}{\partial \ve} d  \bi{k},
\end{align}
where $e$ is the elementary charge ($e > 0$), $f_0$ is the Fermi--Dirac distribution function in thermal equilibrium, and $\Phi$ is the energy variation that depends on the external field. Under an electric field $\bi{E}$ and a magnetic field $\bi{B}$, $\Phi$ can be 
given as
\begin{align}
\label{eq_phi}
\Phi &= 
- e \tau
\left(
\bi{E} \cdot \bi{v} 
+ e \tau \Om [\bi{E} \cdot \bi{v} ]
	\right), \\
\label{ope_om}
\Omega
&=
\frac{1}{\hbar}\sum_{\mu \nu \la} \epsilon_{\mu \nu \la} \frac{\partial \ve}{\partial k_\mu} B _\nu \frac{\partial}{\partial k_\la}  \\
& (\mu, \nu, \la = x,y,z),  \nonumber
\end{align}
Here, $\tau$ and $\epsilon_{\mu \nu \la}$ are the relaxation time and Levi--Civita symbol, respectively. In the low temperature limit, $-\partial f_0 / \partial \ve$ becomes $\delta (\ve - \ve_{\rm F})$, where $\ve_{\rm F}$ is the Fermi energy. For the relativistic carriers, the velocity in Eq. (12) should be replaced by $\bi{v}^{\rm R}$, so that \equref{eq_current} becomes 
\begin{align}
\label{rela_j}
\bi{j}^{\rm R}
      &= 
\frac{e^2 \tau}{(2 \pi)^3}
 \int d \bi{k}  \delta (\ve^{\rm R} -\ve_{\rm F}) \nonumber \\
 &   \times \bi{v}^{\rm R}
\left(
\bi{E} \cdot \bi{v}^{\rm R}
+  e \tau \hat{\Om}^{\rm R} [\bi{E} \cdot  \bi{v}^{\rm R}]
\right) 
  \nonumber \\
      &=    
\la_{\ve_{\rm F}} n e \hat{\mu} \cdot \bi{E} 
 +  \la_{\ve_{\rm F}}^2 n e \hat{\mu} \left[ \bi{B}  \times  (\hat{\mu} \cdot \bi{E}) \right],
\end{align}
where $n$ is the carrier density and $\hat{\mu} = e\tau \hat{\al}$ is the mobility tensor. The current density without magnetic field ($\bi{B}=0$) is represented as $\bi{j}^{\rm R} = \la_{\ve_{\rm F}} ne \hat{\mu} \cdot \bi{E}$.
Thus, the current density under the magnetic field $\bi{j}^{\rm R}$ is obtained:
\begin{align}
\label{eq_current_rela}
\bi{j}^{\rm R}  = 
\la_{\ve_{\rm F}} n e \left[ \hat{ \mu }^{-1} - \la_{\ve_{\rm F}} \hat{B} \right]^{-1}  \cdot \bi{E},
\end{align}
where $\hat{B}$ is represented as the $3 \times 3$ matrix\cite{H.J.Mackey1969,Aubrey1971}:
\begin{align}
\label{eq_mat_B}
\hat{B}
=
\left(
\begin{array}{ccc}
0      & -B_z & B_y   \\
B_z  &   0    & -B_x   \\
-B_y & B_x  & 0  \\
\end{array}
\right).
\end{align}

\begin{figure}[t]
\begin{minipage}[]{85mm}
\centering
      \includegraphics[width = 75mm]{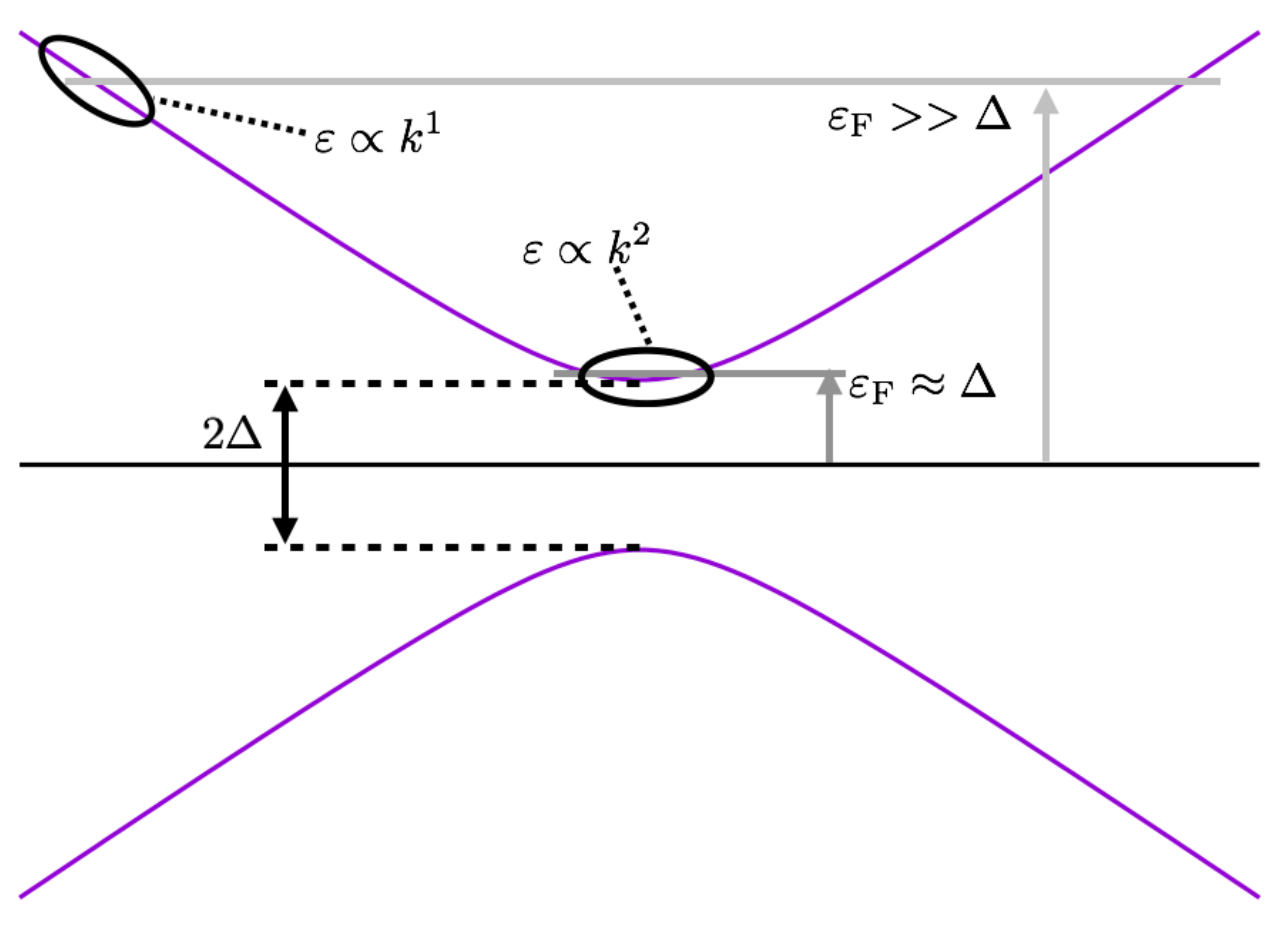}
        \caption{ Relation between the band dispersion of the relativistic electron and the Fermi energy.  \label{fig_band_disp_R_NR}}
\end{minipage}
\end{figure}

Consequently, the relativistic magnetoconducitivity $\hat{\si}^{\rm R}$ is obtained from \equref{eq_current_rela} in the form:
\begin{align}
\label{eq_cdv_rela}
\hat{\si}^{\rm R} &= \la_{\ve_{\rm F}} n e \left[ \hat{\mu} ^{-1} - \la_{\ve_{\rm F}} \hat{B} \right]^{-1}.
\end{align}
This is the  core formula of our work, including one relativistic correction of $\la_{\ve_{\rm F}}$. Note that the non-relativistic magnetoconductivity is described as\cite{H.J.Mackey1969,Aubrey1971,Collaudin2015}
\begin{align}
\label{eq_cdv_nonrela}
\hat{\si}^{\rm NR} &=  n e \left[ \hat{\mu} ^{-1} -  \hat{B} \right]^{-1}.	
\end{align}
Eqs. (\ref{eq_cdv_rela}) and (\ref{eq_cdv_nonrela}) are equal in the limit of $\la_{\ve_{\rm F}} \to 1$ ($\ve_{\rm F} \to \D$), i.e., the so-called non-relativistic limit, where \equref{energy_dirac0} becomes quadratic in $\bi{k}$ as (\figref{fig_band_disp_R_NR})
\begin{align}
\label{eq_Er_nr_approx}
\ve^{\rm R}\to \ve^{\rm NR} = \D + \frac{\hbar^2 }{2} \bi{k} \cdot \hat{\al} \cdot  \bi{k}.
\end{align}
 
\section{Relativistic MR for the one-band model}
In this section, we discuss the relativistic MR for the one-band model and investigate the role of the relativistic correction, $\la_{\ve_{\rm F}}$, on MR. Assuming the isotropic mobility tensor, $\mu_{ij} = \mu_{0} \delta_{ij}$, the elements of \equref{eq_cdv_rela} are
\begin{align}
\label{eq_cdv_rela_xx}
\si_{xx}^{\rm R}     &=      (\mu_0 + \la_{\ve_{\rm F}}  ^2 \eta B_x^2)g^{\rm R}, \\ 
\label{eq_cdv_rela_yy}
\si_{yy}^{\rm R}     &=      (\mu_0 + \la_{\ve_{\rm F}}  ^2 \eta B_y^2)g^{\rm R} ,\\ 
\label{eq_cdv_rela_zz}
\si_{zz}^{\rm R}     &=      (\mu_0 + \la_{\ve_{\rm F}}  ^2 \eta B_z^2)g^{\rm R} ,\\ 
\label{eq_cdv_rela_xy}
\si_{yx}^{\rm R}     &=      (\la_{\ve_{\rm F}}  \mu_0^2 B_z + \la_{\ve_{\rm F}}  ^2 \eta B_y B_x)g^{\rm R} ,\\ 
\label{eq_cdv_rela_yz}
\si_{zy}^{\rm R}     &=      (\la_{\ve_{\rm F}}  \mu_0^2 B_x + \la_{\ve_{\rm F}}  ^2 \eta B_z B_y)g^{\rm R}, \\ 
\label{eq_cdv_rela_zx}
\si_{xz}^{\rm R}     &=      (\la_{\ve_{\rm F}}  \mu_0^2 B_y + \la_{\ve_{\rm F}}  ^2 \eta B_z B_x)g^{\rm R}, \\  
g^{\rm R}     &=    \la_{\ve_{\rm F}}  n e 
\left[ 1 +  \la_{\ve_{\rm F}}  ^2 \mu_0^2 \left(B_x^2 + B_y^2 + B_z^2 \right) \right]^{-1}, \\
\eta     &=     {\rm det}(\hat{\mu}) = \mu_0^3. 
\end{align}

The other elements of $\hat{\si}^{\rm R}$ can be easily obtained from the Onsager relation, $\si_{ij}^{\rm R}(\bi{B}) = \si_{ji}^{\rm R}(-\bi{B})$.  For $\bi{B} = (0,0,B) $, the magnetoresistivity $\rho_{xx}^{\rm R}$ and the Hall resistivity $\rho_{yx}^{\rm R}$ are obtained as follows:
\begin{align}
\label{eq_res_rela_xx}
\rho_{xx}^{\rm R} &= \rho_{yy}^{\rm R}
= \frac{\si_{xx}^{\rm R}}{(\si_{xx}^{\rm R})^2+ (\si_{yx}^{\rm R})^{2}}
=  \frac{1}{ n e \la_{\ve_{\rm F}} \mu_0}, \\
\label{eq_res_rela_yx}
\rho_{yx}^{\rm R} 
&= -\frac{\si_{yx}^{\rm R}}{(\si_{xx}^{\rm R})^2+ (\si_{yx}^{\rm R})^{2}} 
= -\frac{B}{n e} .
\end{align}
One can clearly see the field dependences of $\rho^{\rm R}_{xx}$ and $\rho^{\rm R}_{yx}$, which are what we desired to obtaine. It is found that $\rho_{xx}^{\rm R}$ does not depend on $B$ but its amplitude is modified by the factor of $\la_{\ve_{\rm F}}$ (Fig. 2), whereas $\rho_{yx}^{\rm R}$ is proportional to $B$ but not modified by $\la_{\ve_{\rm F}}$.
The ratios of $\rho_{xx}^{\rm R}$ to $\rho_{xx}^{\rm NR}$ and $\rho_{yx}^{\rm R}$ to $\rho_{yx}^{\rm NR}$ are ($\rho_{ij}^{\rm R}(\la_{\ve_{\rm F}} \to 1) = \rho_{ij}^{\rm NR}$):
\begin{align}
\label{hikaku_res_xx1}
\frac{\rho_{xx}^{\rm R}}{ \rho_{xx}^{\rm NR} } &= \frac{\rho_{yy}^{\rm R}}{ \rho_{yy}^{\rm NR} } = \frac{1}{\la_{\ve_{\rm F}} }, \\
\label{hikaku_res_xy1}
 \frac{\rho_{yx}^{\rm R}}{ \rho_{yx}^{\rm NR}} &= 1.
\end{align}
In addition, the relativistic Hall coefficient $R_{\rm H}^{\rm R}$ is obtained as
\begin{align}
\label{eq_RH_const}
R_{\rm H}^{\rm R}
= -\frac{1}{n e }
\end{align}
for the relativistic electron.
Therefore, the relativistic Hall coefficient $R_{\rm H}^{\rm R}$ is exactly the same as the non-relativistic one $R_{\rm H}^{\rm NR}$. This is consistent with the result for graphene (massless Dirac $\D = 0$)\cite{Peres2007}. 

\begin{figure}[t]
\begin{minipage}[]{87.5mm}
\centering
              \includegraphics[width = 80mm]{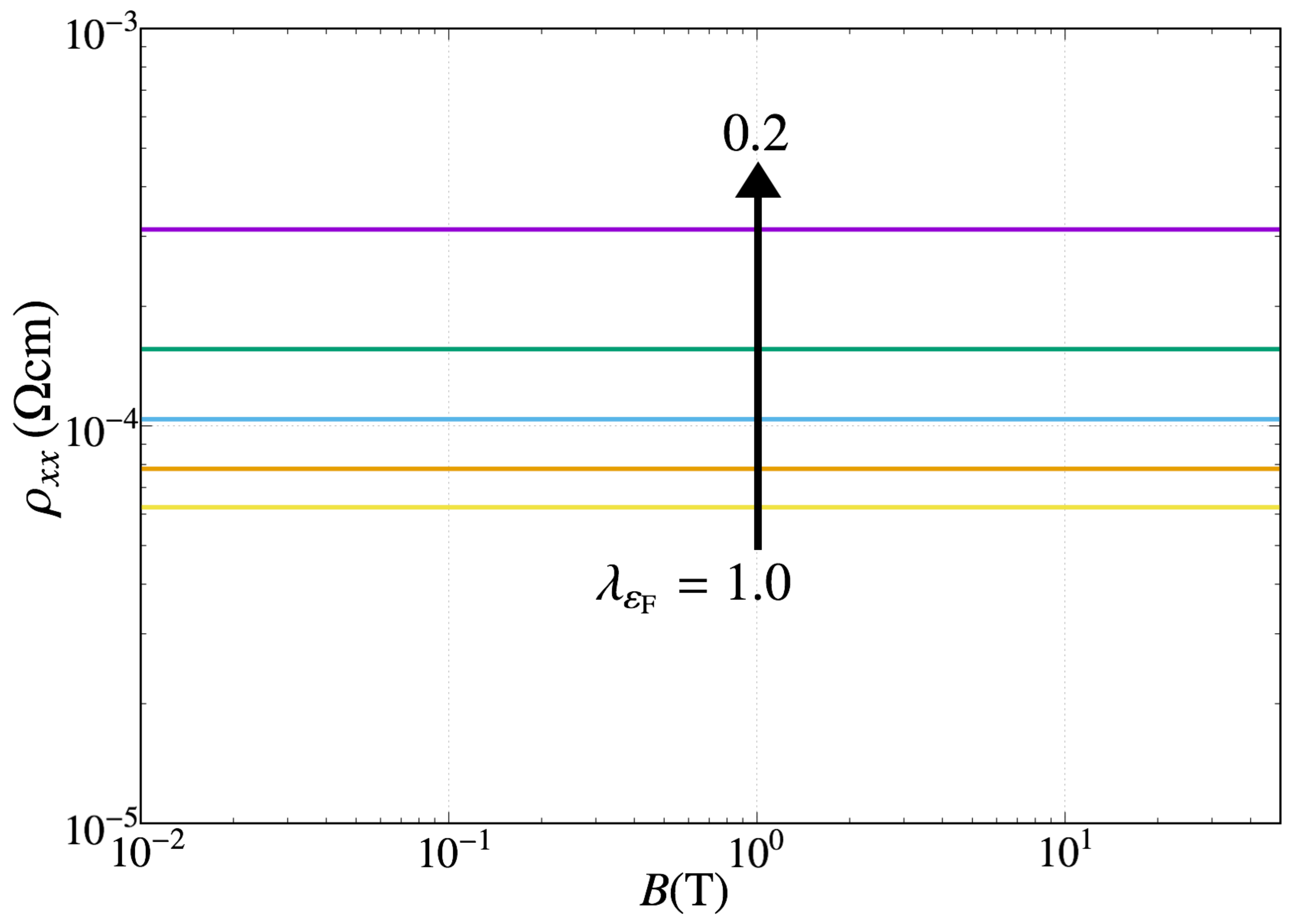}
        \caption{Field dependence of the magnetoresistivity $\rho_{xx}^{\rm R}$ with the different relativistic correction factor $\la_{\ve_{\rm F}} = 0.2, 0.4, 0.6, 0.8, 1.0$. $n$ and $\mu_0$ are set to be $1.0 \times 10^{17} {\rm cm^{-3}}$ and $100{\rm T^{-1}}$, respectively. \label{fig_res1xx}}
\end{minipage}
\end{figure}

\section{Relativistic magnetoresistance for the two-band model}
In the previous section, we discussed the MR for the one-band model. However, the large MR is reported in semimetals. 
Semimetals have electron and hole carriers, so the two-band model is required to explain the transport phenomena in semimetals. In this section, we show the MR of the two-band model. Especially, we discuss the magnetoconductivity in the case with relativistic electron and non-relativistic hole carriers, $\hat{\si}^{\rm R+NR} = \hat{\si}_{\rm e}^{\rm R} + \si_{\rm h}^{\rm NR}$, as in bismuth. We again assume the isotropic mobility tensor for electron and hole carriers $\mu_{ij} = \mu_0 \delta _{ij}$ and $  \nu_{ij} = \nu_0 \delta _{ij}$, respectively. The magnetic field is set to be parallel to the z-axis ($\bi{B} = (0,0,B)$). We also assume that the relativistic correction factor $\la_{\ve_{\rm F}}$ and the number of electrons and holes, $n$ and $p$, do not depend on the magnetic field. (The case where $\la_{\ve_{\rm F}}$, $n$, and $p$ depend on the magnetic field is discussed in Sect. 5.)

\begin{figure*}[t]
\begin{minipage}[]{175mm}
\centering
      \includegraphics[width = 78mm]{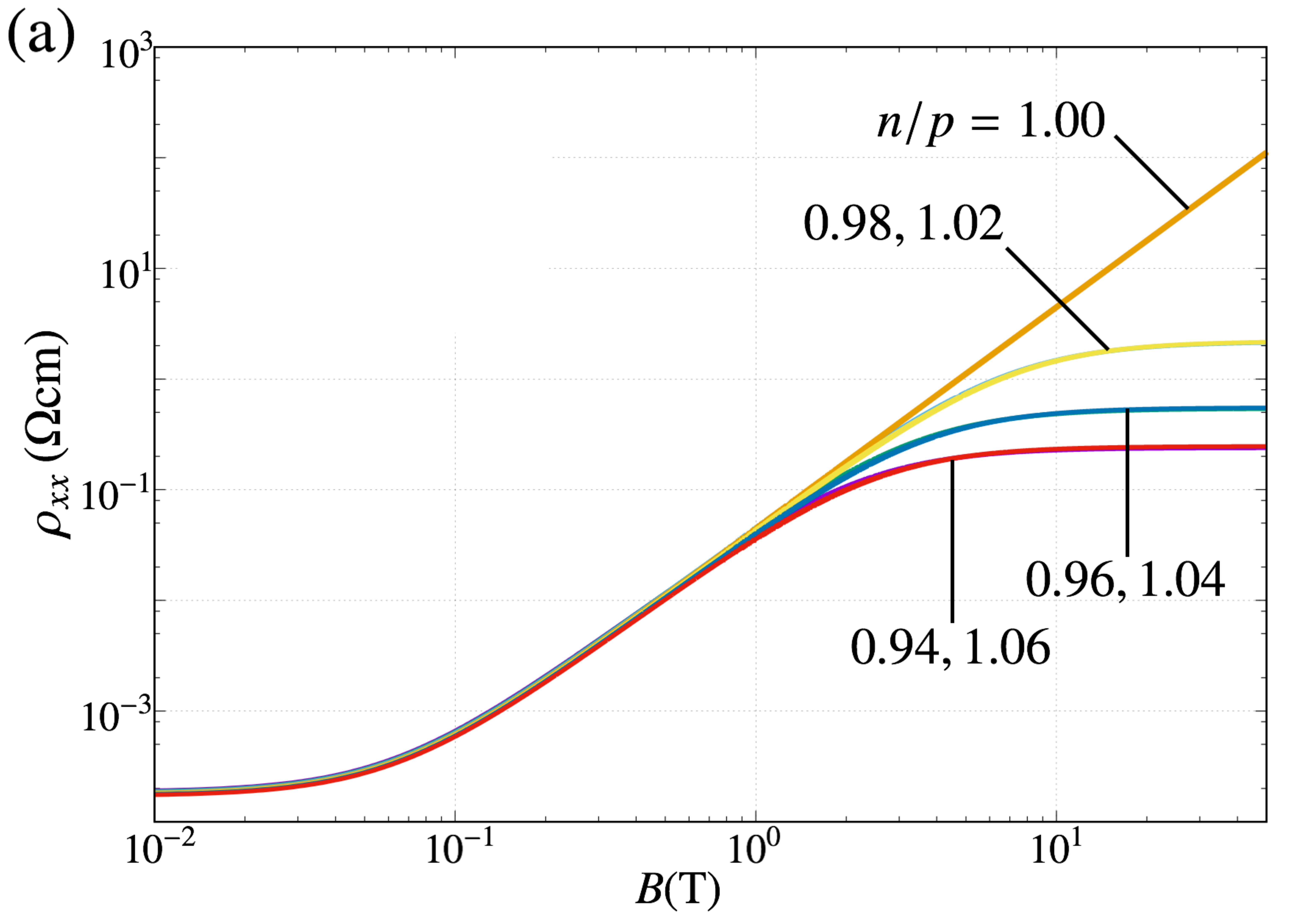} \hspace{30pt}
      \includegraphics[width = 78mm]{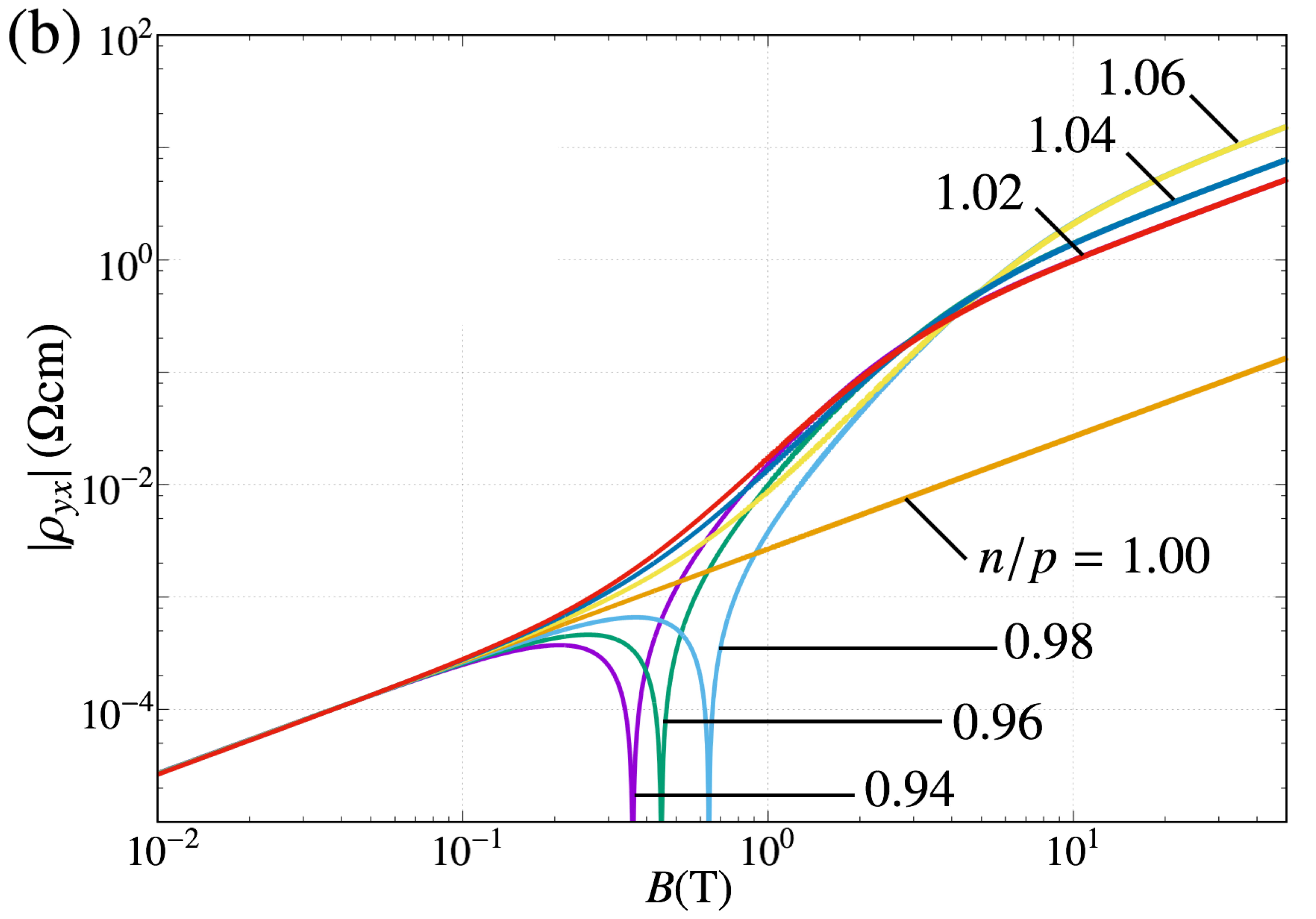}
        \caption{Field dependences of (a) the magnetoresistivity $\rho_{xx}^{\rm R+ NR}$, and (b) the Hall resistivity $\rho_{yx}^{\rm R+ NR}$, for the different electron carriers, $n/p = 0.94, 0.96, 0.98,1.00, 1.02, 1.04, 1.06$ ($p = 1.0 \times 10^{17} \rm cm^{-3}$). $\mu_0, \nu_0$ and $\la_{\ve_{\rm F}}$ are $100 \rm T^{-1}$, $10 \rm T^{-1}$ and 0.25, respectively.  \label{fig_res_npp}}
\end{minipage}
\end{figure*}

The magnetoconductivity $\si_{xx}^{\rm R+NR}$ and the Hall conductivity $\si_{yx}^{\rm R+NR}$ are
\begin{align}
\label{eq_cdv_xx_2}
\si_{xx}^{\rm R+ NR} &=  \si_{yy}^{\rm R+ NR} \nonumber \\
&= \frac{	e\left[ n\la_{\ve_{\rm F}} \mu_0 + p\nu_0 + ( p \la_{\ve_{\rm F}} \mu_0+n\nu_0 )\la_{\ve_{\rm F}} \mu_0 \nu_0 B^2\right]	}{(1+\la_{\ve_{\rm F}}^2 \mu_0^2 B^2)(1+\nu_0^2B^2)},  \nonumber \\ \\
\label{eq_cdv_yx_2}
\si_{yx}^{\rm R+ NR} 
&= 
- \frac{ 	e \left[ (p \nu_0^2 - n\la_{\ve_{\rm F}}^2 \mu_0^2 )B +(p-n) \la_{\ve_{\rm F}}^2 \mu_0^2 \nu_0^2 B^3 \right]	}{(1+\la_{\ve_{\rm F}}^2 \mu_0^2 B^2)(1+\nu_0^2B^2)}.  \nonumber \\
\end{align}
Then, the magnetoresistivity $\rho_{xx}^{\rm R+NR}$ and the Hall resistivity $\rho_{yx}^{\rm R+NR}$ are derived:
\begin{align}
\label{eq_res_xx2}
\rho_{xx}^{\rm R+ NR} &= \rho_{yy}^{\rm R+ NR} \nonumber \\
&=
\frac{1}{e} \frac{	n \la_{\ve_{\rm F}}\mu_0 + p\nu_0 + \la_{\ve_{\rm F}}\mu_0\nu_0 B^2(n\nu_0 + p \la_{\ve_{\rm F}}\mu_0 )	}{	(n\la_{\ve_{\rm F}}\mu_0 + p \nu_0 )^2 + \la_{\ve_{\rm F}}^2\mu_0^2 \nu_0^2 B^2 (p-n)^2	}, \nonumber \\ \\
\label{eq_res_yx2}
\rho_{yx}^{\rm R+ NR} &= -\rho_{xy}^{\rm R+ NR} \nonumber \\
&= 
\frac{1}{e} \frac{		(p\nu_0^2 - n \la_{\ve_{\rm F}}^2\mu_0^2 )B + \la_{\ve_{\rm F}}^2\mu_0^2	 \nu_0^2 B^3 (p-n)}{		(n\la_{\ve_{\rm F}}\mu_0 + p \nu_0 )^2 + \la_{\ve_{\rm F}}^2\mu_0^2 \nu_0^2 B^2 (p-n)^2		}.
\nonumber \\ 
\end{align}
Again, we obtain the analytic forms of $\rho_{xx}(B)$ and $\rho_{yx}(B)$, where their field dependences are clearly indicated. Figure \ref{fig_res_npp} shows the magnetic field dependence of $\rho_{xx}^{\rm R+NR}$ and $\rho_{yx}^{\rm R+NR}$ for $p = 1.0 \times 10^{17}{\rm cm^{-3}},\mu_0 = 100{\rm T^{-1}}, \nu_0 = 10{\rm T^{-1}}$, and $\la_{\ve_{\rm F}} = 0.25$, which are consistent with bismuth for $B\parallel$ trigoal axis. The lines on both figures have different values of $n/p = 0.94 \sim  1.06$.
$\rho_{xx}^{\rm R+NR}$ has a constant value under weak fields ($B \to 0 \rm T$). $\rho_{xx}^{\rm R+NR}$ increases as $\rho_{xx}^{\rm R+NR} \propto B^2$, and saturates under strong fields except for $n/p = 1.00$. The saturated value becomes small when the difference between $n$ and $p$ becomes large. It is seen from \figref{fig_res_npp}(b) that the sign of $\rho_{yx}^{\rm R+NR}$ with $n/p < 1.00$ changes at a certain magnetic field. $|\rho_{yx}^{\rm R+NR}|$ is the same under strong fields when $\rho_{yx}^{\rm R+NR}$ has the same difference from $n/p = 1.00$ (e.g., $n/p = 0.98, 1.02$).

The behaviors of $\rho_{xx}(B)$ and $\rho_{yx}(B)$ can be understood more clearly if we take the limit of weak and strong fields. At weak fields ($\la_{\ve_{\rm F}}^2 \mu_0^2 B^2 \ll 1, \nu_0^2B^2 \ll 1$), \equsref{eq_res_xx2} and (\ref{eq_res_yx2}) become 
\begin{align}
\label{eq_res_xx2_w}
\rho_{xx}^{\rm R+ NR}
&= \frac{1}{e}\frac{1}{(n  \la_{\ve_{\rm F}} \mu_0 + p\nu_0)}, \\
\label{eq_res_xy2_w}
\rho_{yx}^{\rm R+ NR} 
&= \frac{1}{e} \frac{\left( p- n \la_{\ve_{\rm F}}^2 \ka^2 \right)B}{(p +n \la_{\ve_{\rm F}} \ka )^2},
\end{align}
where
\begin{align}
\ka &= \frac{\mu_0}{\nu_0}. 
\end{align}
expresses the difference between electron and hole mobilities.
From \equref{eq_res_xy2_w}, the Hall coefficient $R_{\rm H}^{\rm R+NR}$ is obtained:
\begin{align}
\label{eq_hall_coeff_RNR}
R _{\rm H}^{\rm R + NR}
&= \frac{1}{e} \frac{p - \la_{\ve_{\rm F} }^2n  \ka ^2}{ \left[ p + \la_{\ve_{\rm F}} n  \ka \right]^2}. 
\end{align}
Equation (\ref{eq_hall_coeff_RNR}) is similar to the Hall coefficient of a system with non-relativistic electrons and holes, which is given by \cite{Kittel8th}:
\begin{align}
\label{eq_hall_coeff_NRNR}
R _{\rm H}^{\rm NR + NR}
=\frac{1}{e} \frac{p - n  \ka ^2}{ \left[ p + n  \ka \right]^2}.
\end{align}
An important finding here is that the Hall coefficient $R_{\rm H}^{\rm R+NR}$ includes the relativistic correction factor $\la_{\ve_{\rm F}}$, while $R_{\rm H}^{\rm R}$ does not for the one-band model. 

At strong fields ($\la_{\ve_{\rm F}}^2  \mu_0^2 B^2, \nu_0^2 B^2 \gg 1$), \equsref{eq_res_xx2} and (\ref{eq_res_yx2}) can be represented as follows: 
\begin{align}
\label{eq_res_xx2_st}
\rho_{xx}^{\rm R+ NR} 
=\left\{ \begin{array}{ll}
\displaystyle \frac{1}{ne} \frac{\la_{\ve_{\rm F}} \mu_0 \nu_0 B^2}{\nu_0 + \la_{\ve_{\rm F}}\mu_0} & (n=p) \vspace{5pt}\\
\displaystyle \frac{n \nu_0 + p \la_{\ve_{\rm F}} \mu_0}{ (n-p)^2 e \la_{\ve_{\rm F}} \mu_0 \nu_0} & (n \neq p) \\
\end{array} \right.  \\
\label{eq_res_yx2_st}
\rho_{yx}^{\rm R+ NR} 
=\left\{ \begin{array}{ll}
\displaystyle \frac{1}{ne} \frac{(\nu_0 - \la_{\ve_{\rm F}} \mu_0) B}{\nu_0 + \la_{\ve_{\rm F}}\mu_0} & (n=p) \vspace{5pt}\\
\displaystyle -\frac{B}{ (n-p)e} & (n \neq p) \\
\end{array} \right. 
\end{align}
From \equref{eq_res_xx2_st}, $\rho_{xx}^{\rm R+NR}$ is saturated, when $n \neq p$. In contrast, $\rho_{xx}^{\rm R+NR}$ increases as $\rho_{xx}^{\rm R+NR} \propto B^2$ when $n$ and $p$ are completely equal ($n=p$). Despite the fact that relativistic and non-relativistic electrons have different band dispersions, $\rho_{xx}^{\rm R+NR}$ and $\rho_{xx}^{\rm NR+NR}$ increase with $B^2$ for $n = p$, which is very interesting.  The relativistic effect does not qualitatively change the magnetic field dependence of $\rho_{xx}$.
Note that, the formula $\rho_{yx}^{\rm R+NR}$ is the same as $\rho_{yx}^{\rm NR+NR}$ when $n \neq p$\cite{Kittel8th}. The $\la_{\ve_{\rm F}}$ dependences of the magnetoresistivities ($\rho_{ij}^{\rm R}$, $\rho_{ij }^{\rm R+NR}$) and the Hall coefficient $R_{\rm H}^{\rm R, R+NR}$ are summarized in \tabref{tab_la_B_res}. It is apparent that $\la_{\ve_{\rm F}}$ dependence of $\rho_{xx}^{\rm R+NR}$ is different at weak fields and strong fields. 

\begin{figure*}[p]
\begin{minipage}[]{170mm}
\centering
	 \includegraphics[width = 75mm]{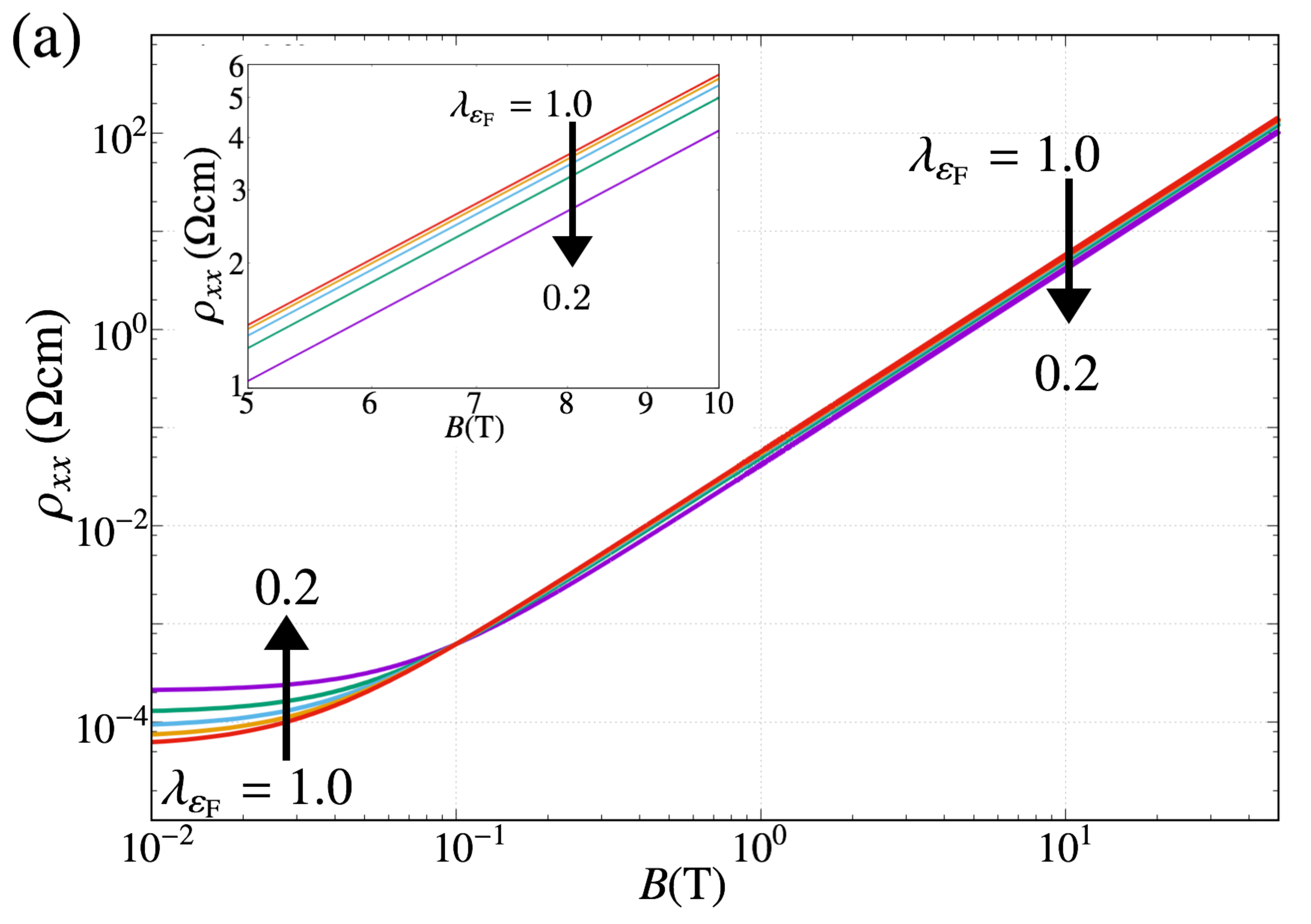}  \hspace{30pt}
	 \includegraphics[width = 75mm]{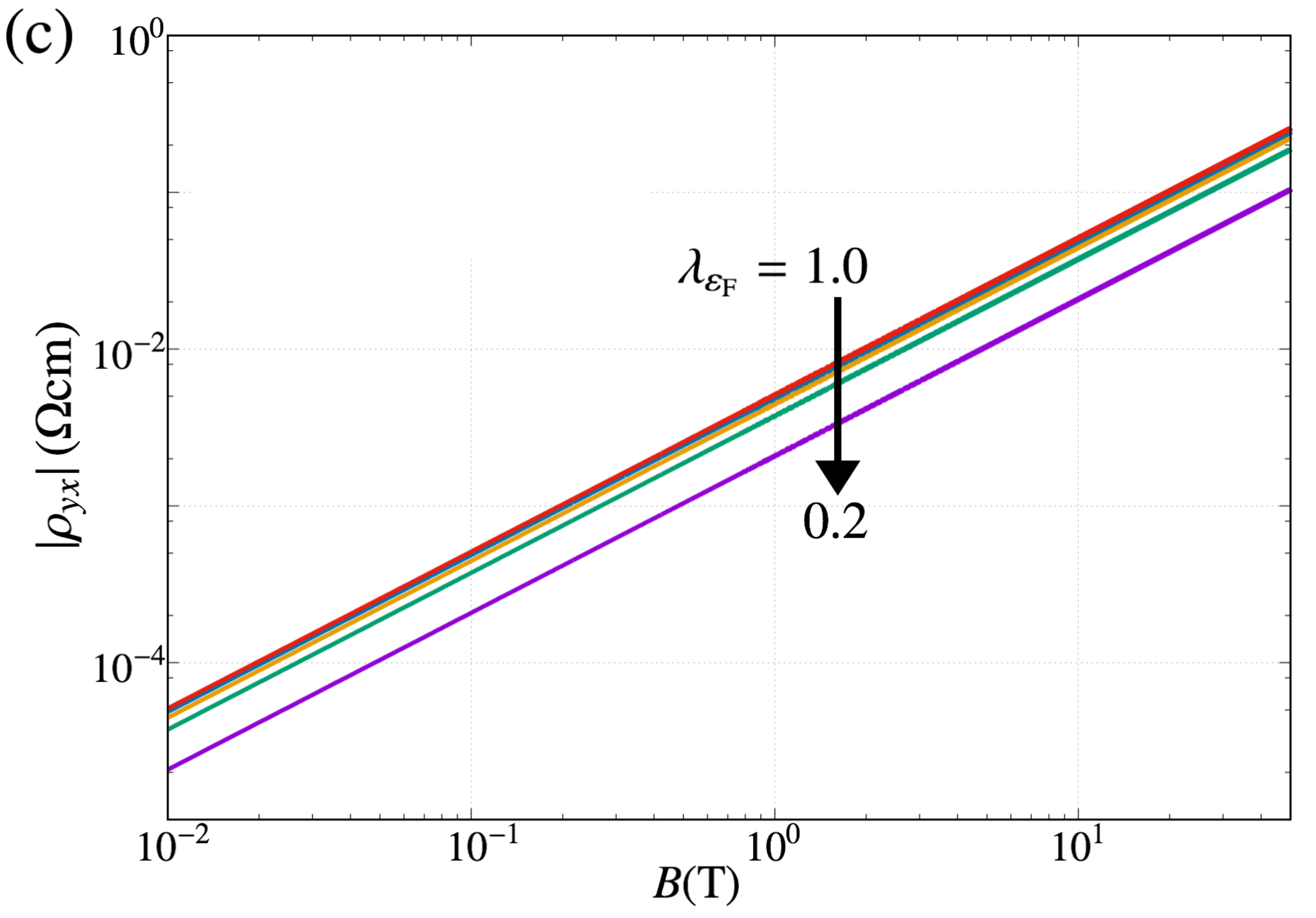} \vspace{5pt} \\
	 \includegraphics[width = 75mm]{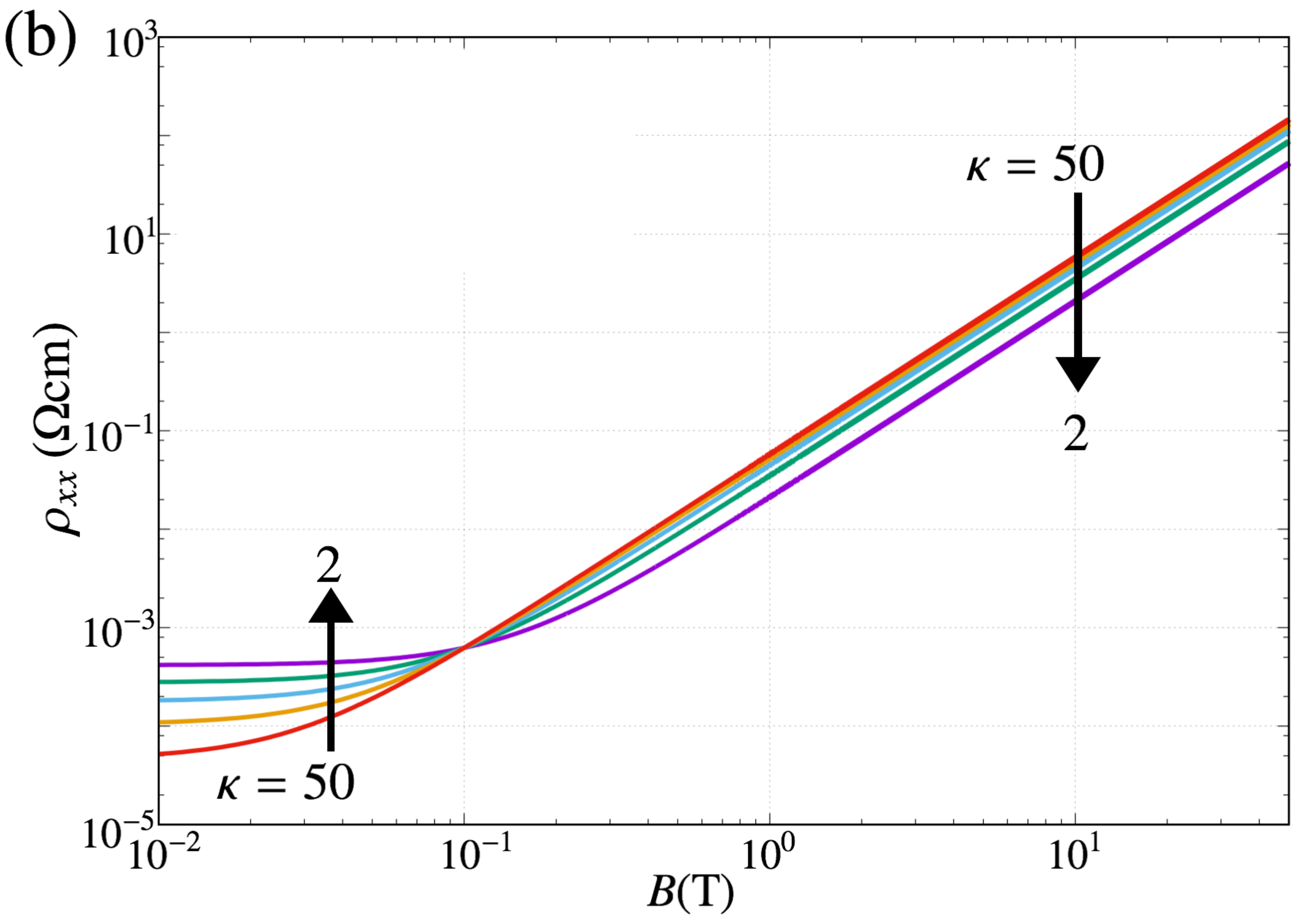}\hspace{30pt}
	 \includegraphics[width = 75mm]{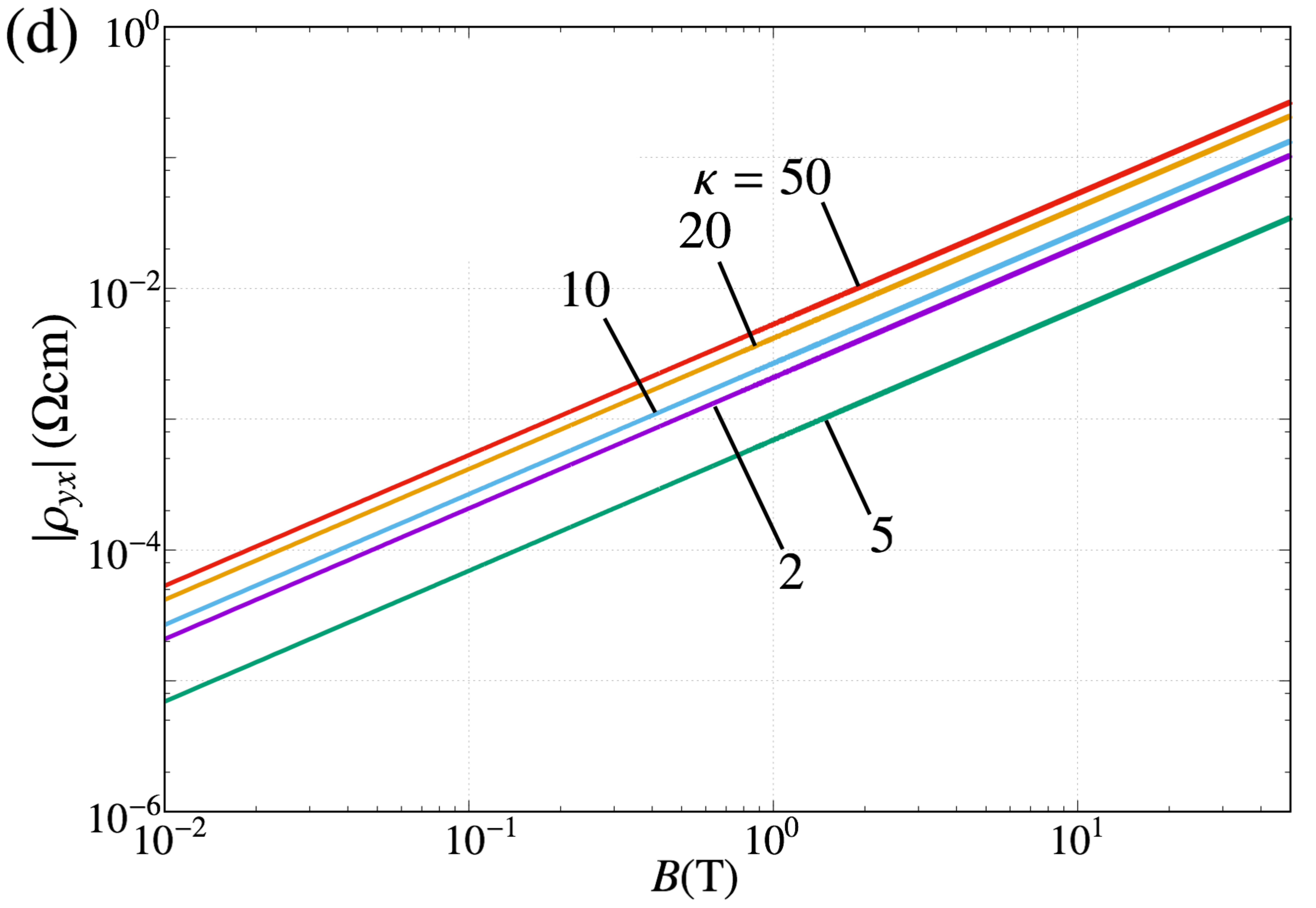}
	         \caption{Field dependence of (left) the magnetoresistivity $\rho_{xx}^{\rm R+ NR}$, and (right) The Hall resistivity $\rho_{yx}^{\rm R+ NR}$ for (top) the different relativistic correction factors, $\la_{\ve_{\rm F}} = 0.2, 0.4, 0.6, 0.8, 1.0$, with $\mu_0 = 100 \rm T^{-1}$ and $\nu_0 = 10 \rm T^{-1}$, and (bottom) the different ratio of $\mu_0$ to $\nu_0$, $\ka = 2, 5, 10, 20, 50$, with $\la_{\ve_{\rm F}} = 0.25$ and $\nu_0 = 10 \rm T^{-1}$. The electron and hole carriers ($n,p$) are set to be $1.0 \times 10^{17} \rm cm^{-3}$ on (a)-(d). \label{fig_res_L_ka}}
\end{minipage} 
\end{figure*}

In the following, the $\la_{\ve_{\rm F}}$ and $\ka $ dependence of $\rho_{xx}^{\rm R+NR}$ and $\rho_{yx}^{\rm R+NR}$ are evaluated for $n = p = 1.0 \times 10^{17} \rm cm^{-3}$. The lines in \figsref{fig_res_L_ka}(a) and (c) have different $\la_{\ve_{\rm F}}$ ($\la_{\ve_{\rm F}} = 0.2, 0.4, 0.6, 0.8, 1.0$). $\rho_{xx}^{\rm R+NR}$ increases as $\la_{\ve_{\rm F}}$ decreases at weak fields, whereas $\rho_{xx}^{\rm R+NR}$ decreases with $\la_{\ve_{\rm F}}$ at strong fields. $\rho_{yx}^{\rm R+NR}$ decreases as $\la_{\ve_{\rm F}}$ decreases.
\begin{figure*}[p]
\begin{minipage}[]{170mm}
\centering
	 \includegraphics[width = 75mm]{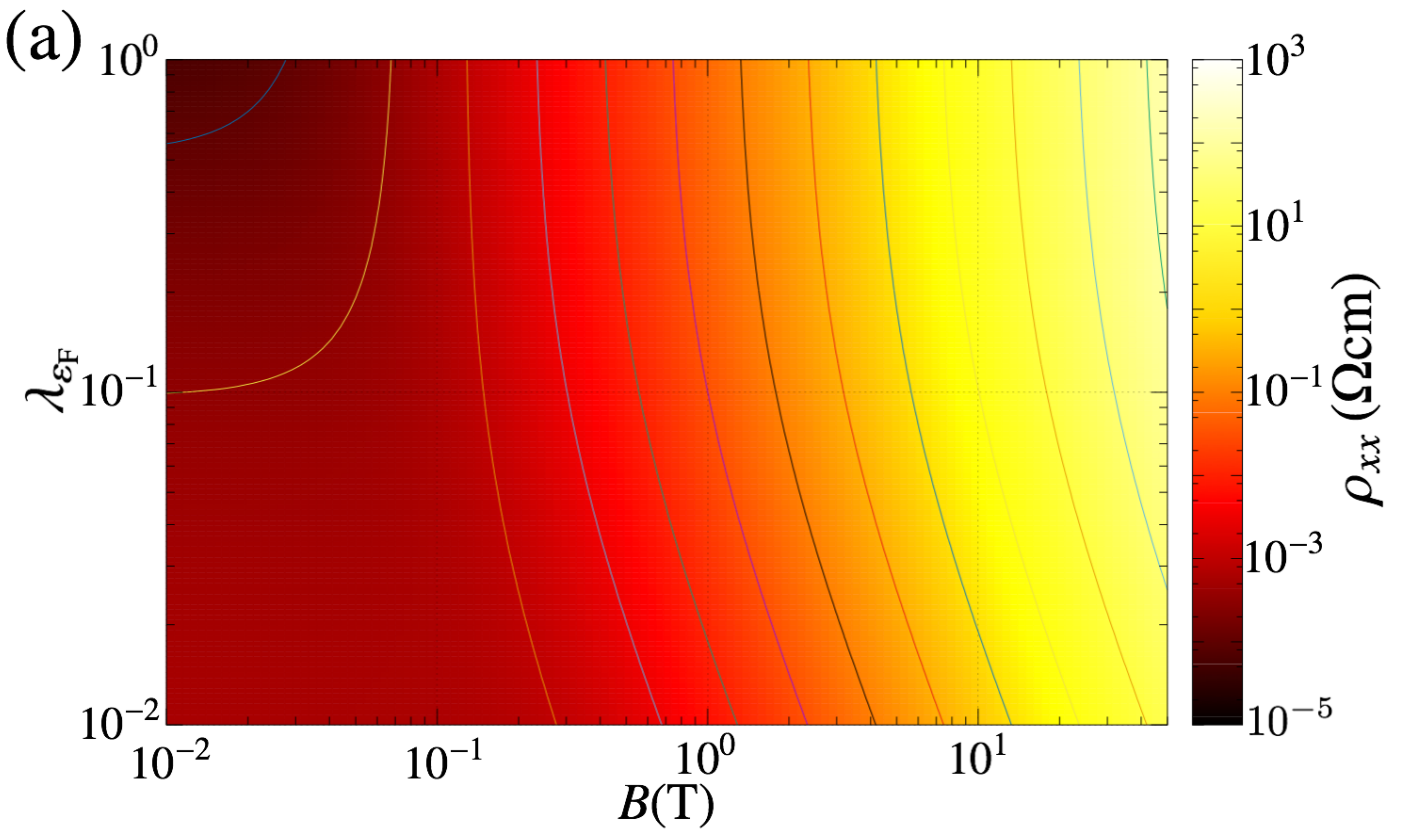}\hspace{30pt}
	 \includegraphics[width = 75mm]{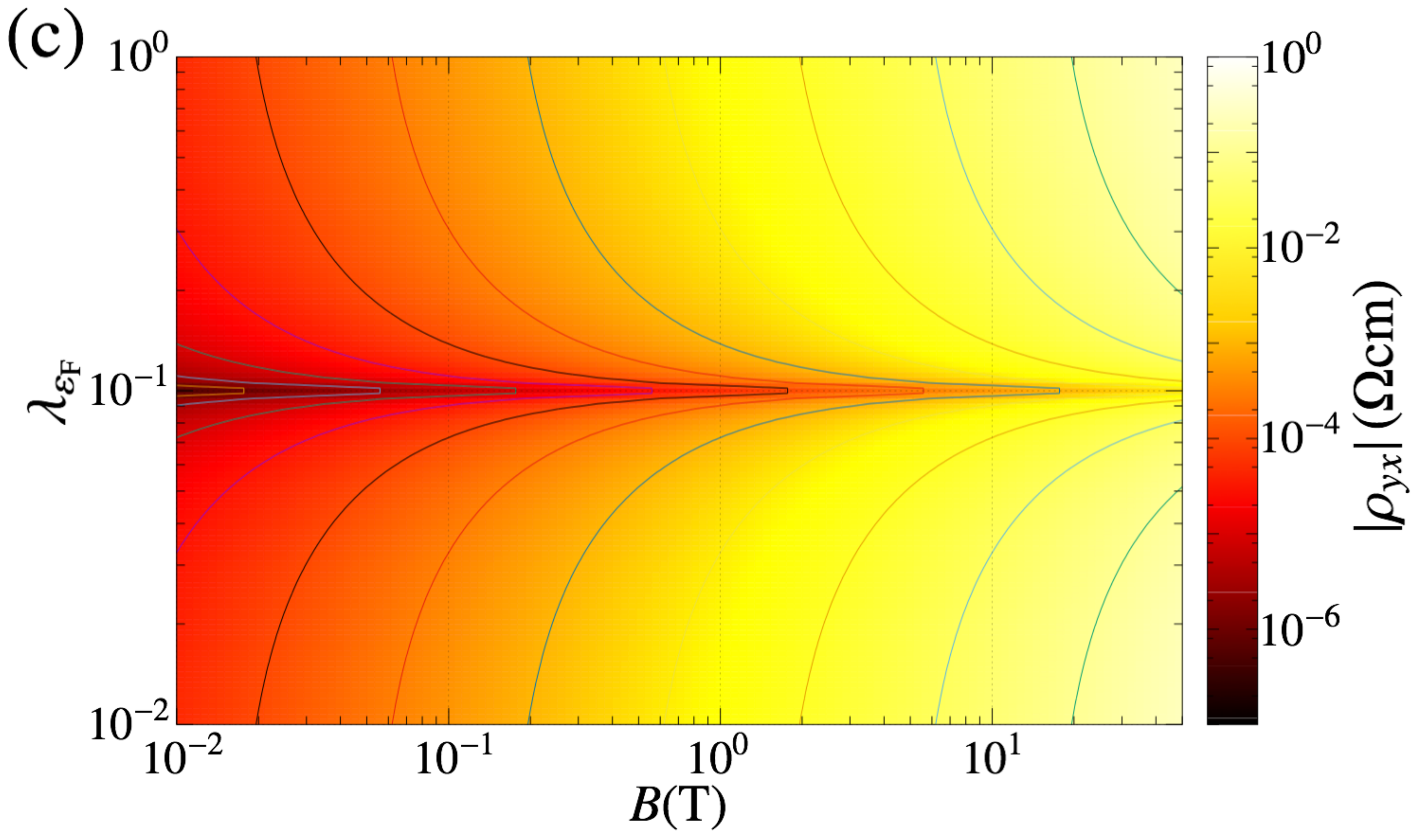}\vspace{5pt}\\
	 \includegraphics[width = 75mm]{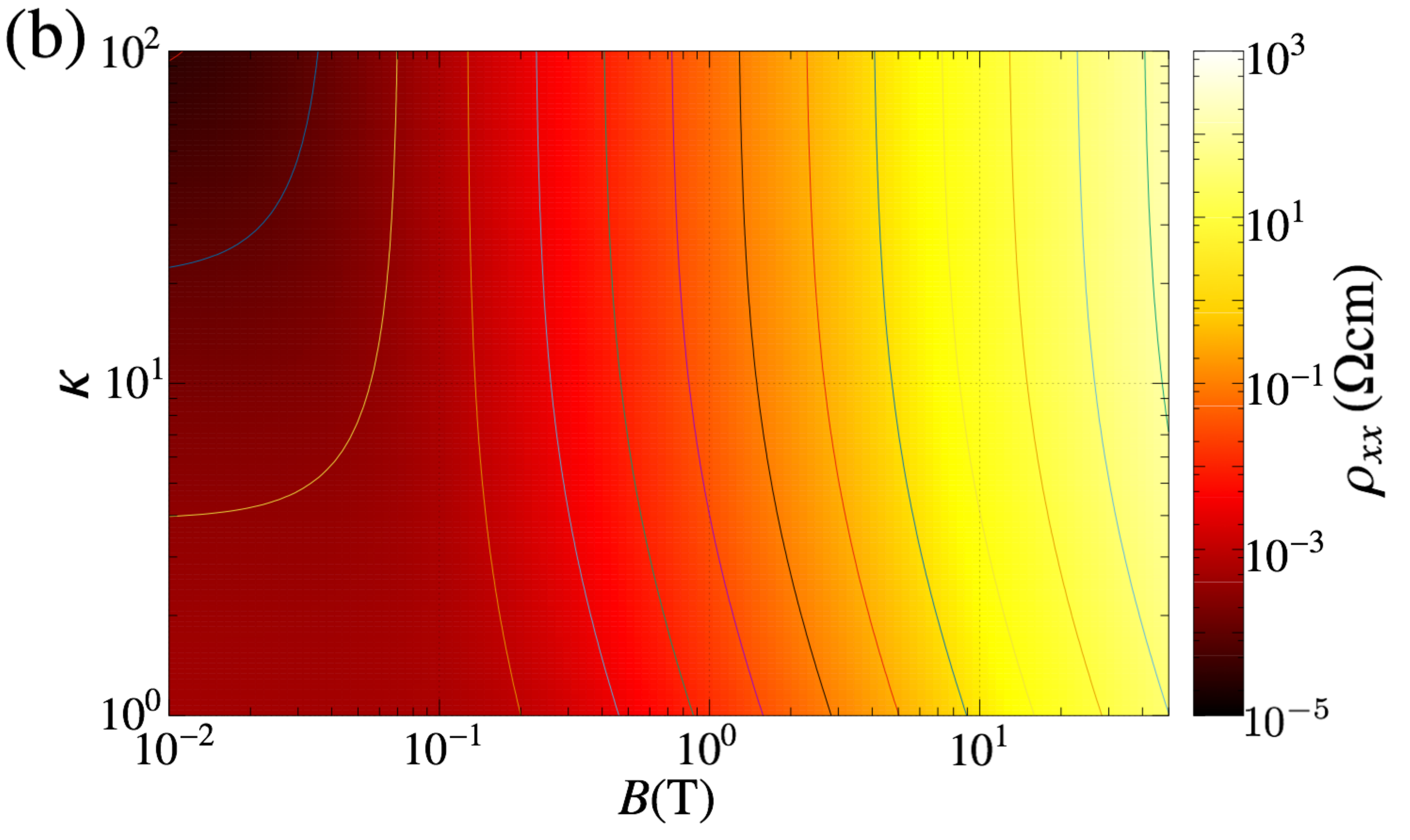}\hspace{30pt}
	 \includegraphics[width = 75mm]{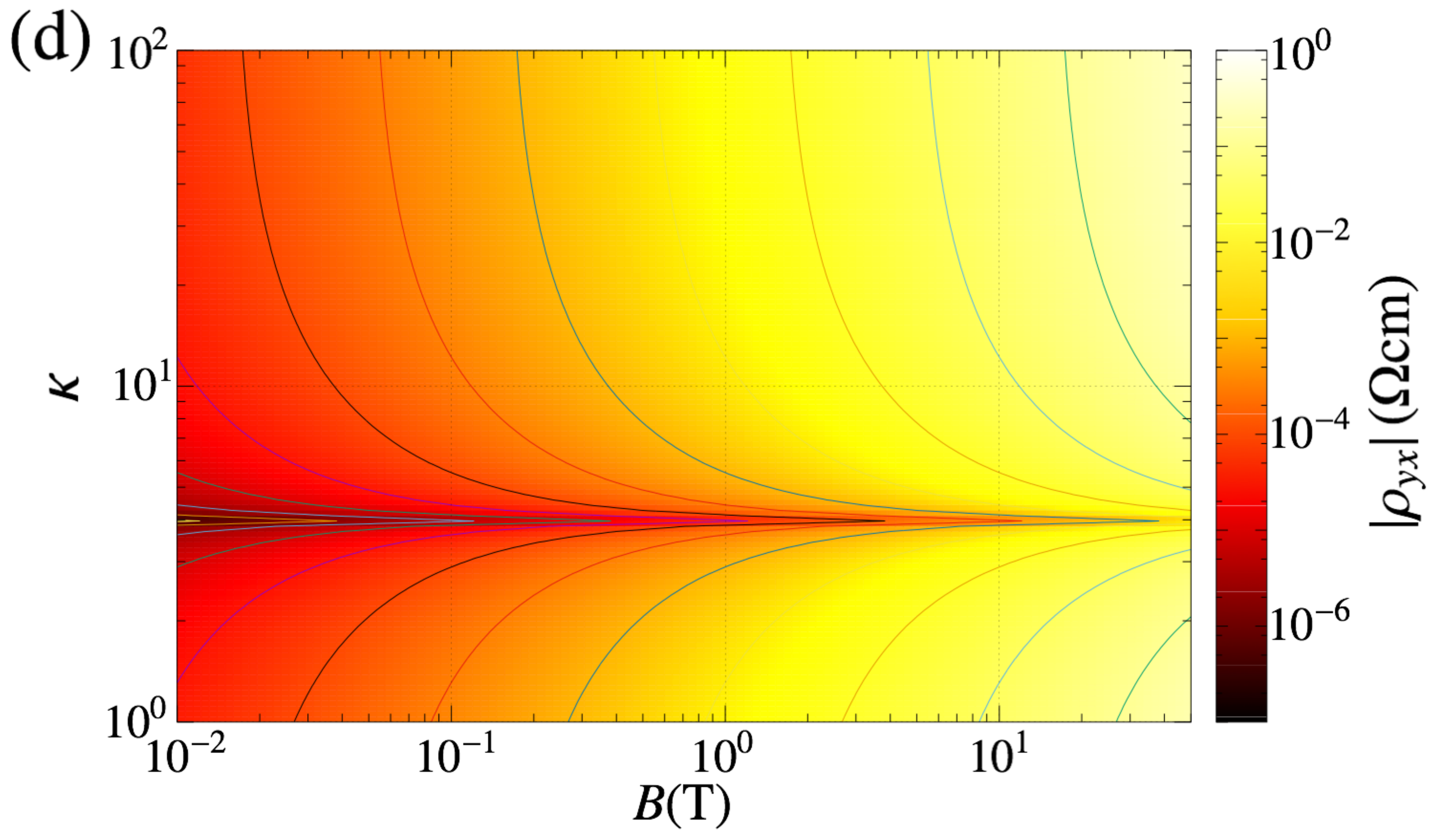}
	         \caption{Color maps of (a) and (b) magnetoresistivity $\rho_{xx}^{\rm R+NR}$, (c) and (d) the Hall resistivity $\rho_{yx}^{\rm R+NR}$. The maps (a) and (c) are with shifted $B$ and $\la_{\ve_{\rm F}}$ at $\ka = 10$, (b) and (d) are with shifted $B$ and $\ka$ at $\la_{\ve_{\rm F}}=0.25$. On the all figures, $\nu_0$ is $10{\rm T^{-1}}$, the electron and hole carriers ($n, p$) are set to be $1.0 \times 10^{17} \rm cm^{-3}$. \label{fig_res_map}}
\end{minipage}
 \end{figure*}
Figures \ref{fig_res_L_ka}(b) and (d) show the field dependence of $\rho_{xx}^{\rm R+NR}$ and $\rho_{yx}^{\rm R+NR}$ with different $\ka$ ($\ka = 2, 5, 10, 20, 50$). Comparing \figref{fig_res_L_ka}(a) with \figref{fig_res_L_ka}(b), it is evident that their effects are similar. $\la_{\ve_{\rm F}}$ and $\mu_0$ always appear in the form of $\la_{\ve_{\rm F}}\mu_0$ in \equsref{eq_res_xx2_w} and (\ref{eq_res_xx2_st}), thus, $\la_{\ve_{\rm F}}$ and $\mu_0$ give the same contribution to $\rho_{xx}^{\rm R+NR}$. In fact, \figsref{fig_res_map}(a) and \ref{fig_res_map}(b) show the same variation of $\rho_{xx}^{\rm R+NR}$. The $\la_{\ve_{\rm F}}$ and $\ka$ dependence of $\rho_{xx}^{\rm R+NR}$ exhibit the same variation. Figures \ref{fig_res_map}(c) and (d) also have the same variation of $\rho_{yx}^{\rm R+NR}$. The lines on \figsref{fig_res_L_ka}(c) and \ref{fig_res_L_ka}(d) have different dependences for $\la_{\ve_{\rm F}}$ and $\ka$. However, this is caused by using $\la_{\ve_{\rm F}} >  0.1$. Note that the position of the fixed point (or cross point) field is determined by $\nu_0$, so that it is $B=0.1 {\rm T}$ in the case with $n = p = 1.0 \times 10^{17} \rm cm^{-3}$, $\nu_0 = 10 \rm T^{-1}$ as is shown in \figsref{fig_res_L_ka}(a) and (b). 

\section{Near the quantum limit}
At weak magnetic fields, the Fermi energy $\ve_{\rm F}$ and carrier densities $n$ and $p$ do not change in three-dimensional systems. This is because carrier energy is not quantized clearly. At strong fields, on the other hand, the electron energy is clearly quantized into the Landau levels, and $\ve_{\rm F}$, $n$ and $p$ of semimetals drastically change with magnetic field in order to keep the charge neutrality. This tendency becomes more significant when the difference between electron and hole mobilities becomes large, such as in bismuth. In this section, we discuss the MRs for the one-band and the two-band model near the quantum limit (QL), where all carriers occupy the lowest Landau level only.
\subsection{Relativistic one-band model near the QL }
To calculate magnetoresistivities $\rho_{xx}^{\rm R}$ and $\rho_{yx}^{\rm R}$ near the QL more accurately, it is necessary to calculate the eigenenergies of each Landau level.  The eigenvalue of Wolff model under magnetic fields is \cite{Wolff1964,Fuseya2015a}
\begin{align}
\label{eq_wolff_mag}
\ve^{\rm R} 
& = 
 \sqrt{\D^2 + 2\D \left[ \left( l + \frac{1}{2} +  \frac{\si}{2} \right)  \hbar \om_{c}+ \frac{\hbar ^2 k_z^2}{2m_z} \right]}, 
\end{align}
where $l$ is an index of the Landau levels, $\si $ is the sign of the spin, $k_z$ is the wavenumber parallel to the magnetic field, $\om_{\rm c} = \beta_0 B/m_{\rm c} $ ($m_c$ is the cyclotron mass, $\beta_0$ is the double amount of the Bohr magneton $\mu_{\rm B}$), $m_z$ is the effective mass for electrons parallel to the magnetic field. In the following, we set $\D = 7.5 \rm meV$, $m_{\rm c}/m_0 = m_z/m_0 = 0.01$.

\begin{figure*}[t]
\begin{minipage}[]{170mm}
\centering
\includegraphics[width = 75mm]{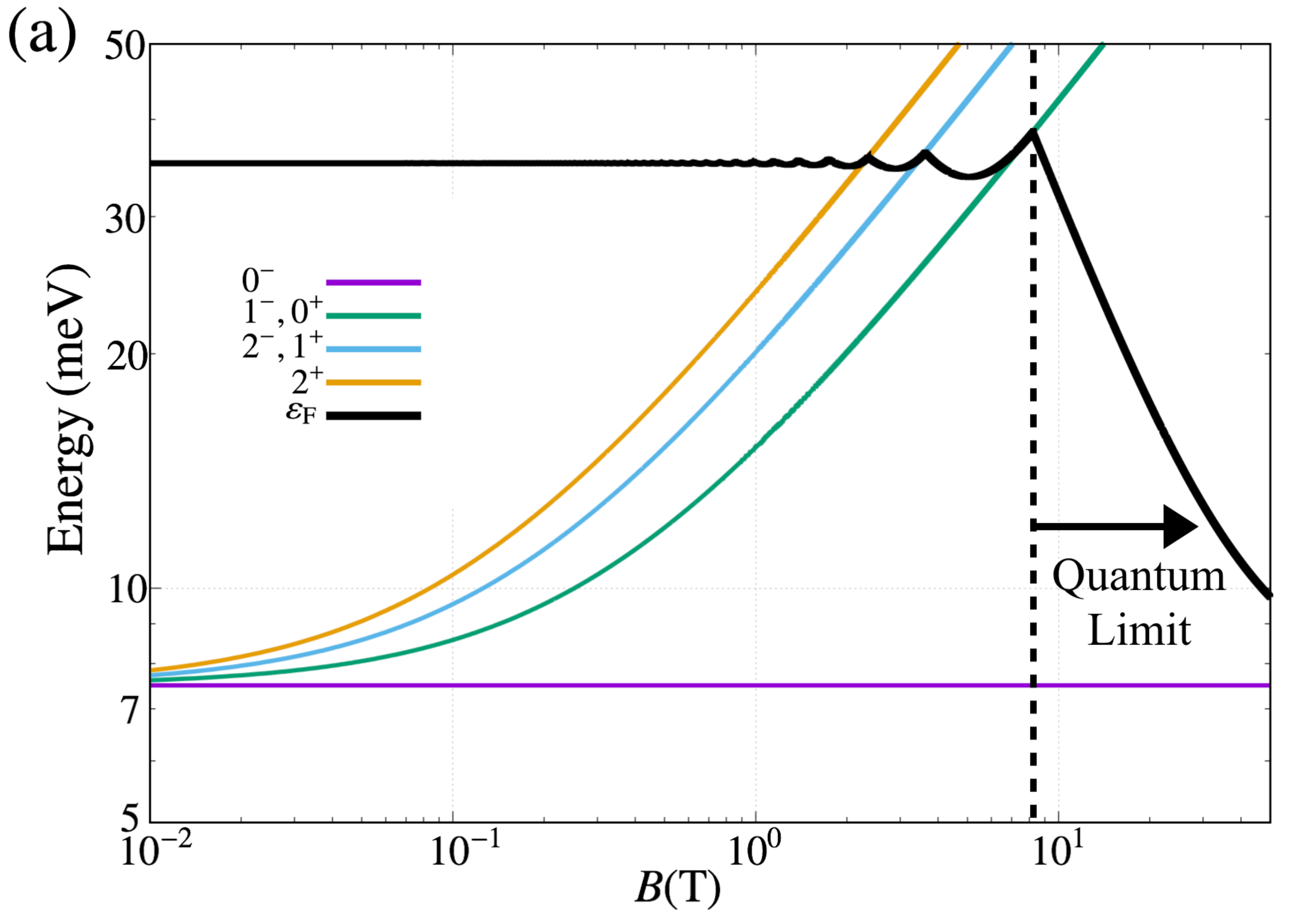} \hspace{30pt}
\includegraphics[width = 75mm]{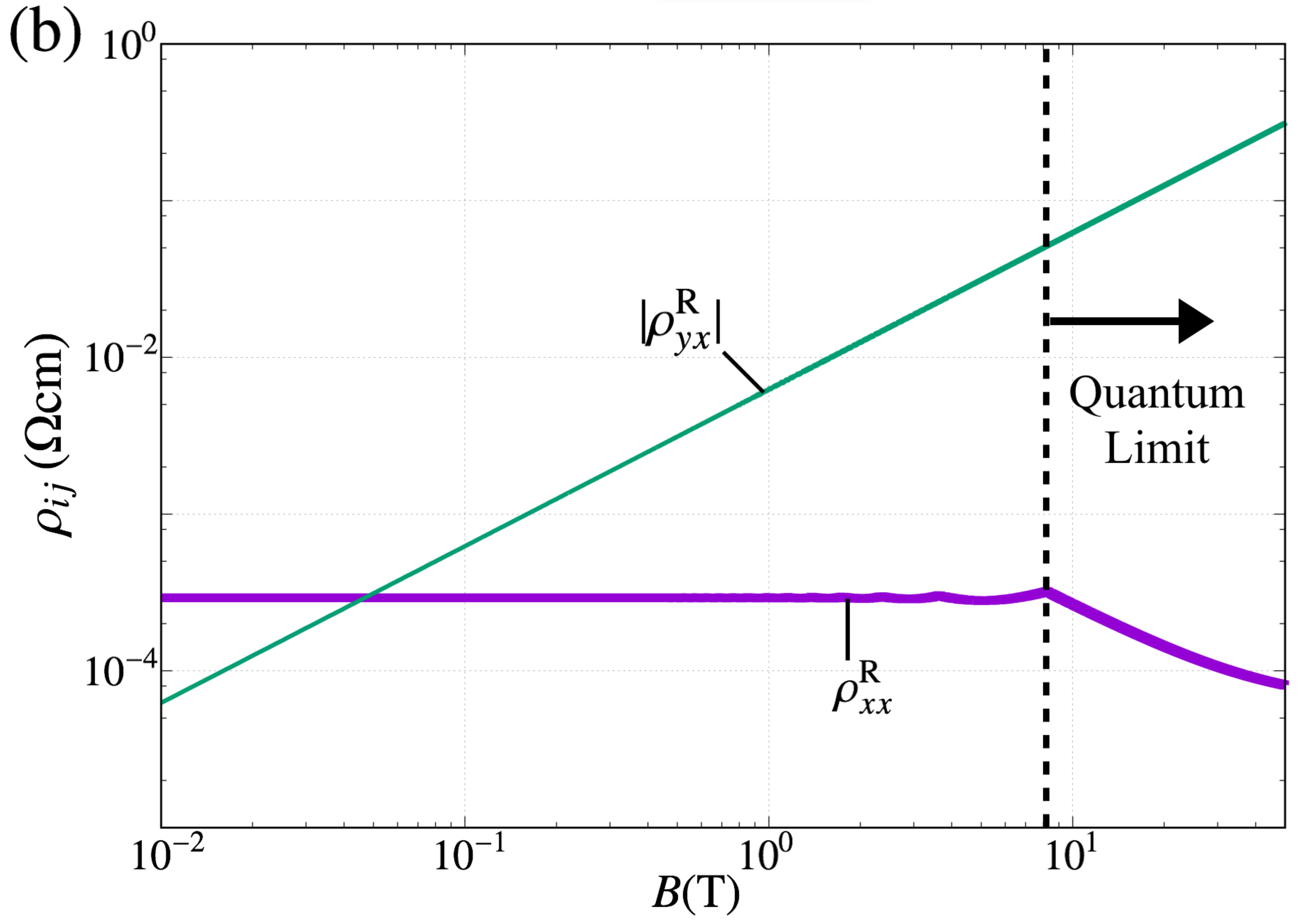}
        \caption{Field dependence of (a) the energies, and (b) the resistivities $\rho_{ij}^{\rm R}$. $l^{\pm}$ on (a) is the Landau index. The thick line and the thin line on (b) show $\rho_{xx}^{\rm R}$ and $|\rho_{yx}^{\rm R}|$, respectively. The electron carrier $n$ is $1.0 \times 10^{17} \rm cm^{-3}$, and the mobility $\mu_0$ is $100 \rm T^{-1}$. \label{fig_r_ql}}
\end{minipage}
\end{figure*}

The formula of carrier density is represented as 
\begin{align}
\label{eq_carrier}
n= \frac{eB}{2\pi^2} \sum_{l\si} \hbar k_{l \si}.
\end{align}
All carriers occupy the lowest Landau level under the QL $(l, \si) = (0,-1)$, so \equref{eq_carrier} is changed as
\begin{align}
\label{eq_ql_pf}
k_{\rm F} = \frac{2 \pi^2 }{\hbar eB} n, 
\end{align}
where $k_{\rm F}$ is the Fermi wavenumber. 
The charge is conserved ($n$ = const.), thus, $k_{\rm F} \propto B^{-1}$. This implies that $\ve_{\rm F}$ depends on the magnetic field (\figref{fig_r_ql}(a)).
With the present parameters, electrons reach the QL at $B \simeq 8$T and the Fermi energy $\ve_{\rm F}$ moves down to the band edge ($\ve_{\rm F} \rightarrow \D$) beyond this field. The behavior of $\rho_{xx}^{\rm R}$ and $\rho_{yx}^{\rm R}$ is shown in \figref{fig_r_ql}(b).
The field dependence of $\rho_{yx}^{\rm R}$ does not change in the QL region, because it does not depend on $\la_{\ve_{\rm F}}$. In contrast, $\rho_{xx}^{\rm R}$ attains field dependence under the QL because $\rho_{xx} \propto \la_{\ve_{\rm F}}^{-1} = \ve_{\rm F}/ \D$. $\la_{\ve_{\rm F}}$ is close to unity (non-relativistic limit) as the Fermi energy $\ve_{\rm F}$ moves down to $\D$ under the QL. Therefore, even if the electrons are relativistic under weak fields, they exhibit the non-relativistic behavior under the QL.
\subsection{Two-band model (R+NR) near the QL}
In the two-band model, the variation of the Fermi energy is more complicated. For example, $\ve_{\rm F}$ in bismuth moves downward with the increase in magnetic field parallel to the binary axis and moves upward with the fields parallel to the trigonal axis\cite{Zhu2011PRB}. The former makes $\la_{\ve_{\rm F}}$ large and the latter makes $\la_{\ve_{\rm F}}$ small. Here, we discuss the system where $\ve_{\rm F}$ moves downward.

\begin{figure*}[t]
\begin{minipage}[]{170mm}
\centering
\includegraphics[width = 73mm]{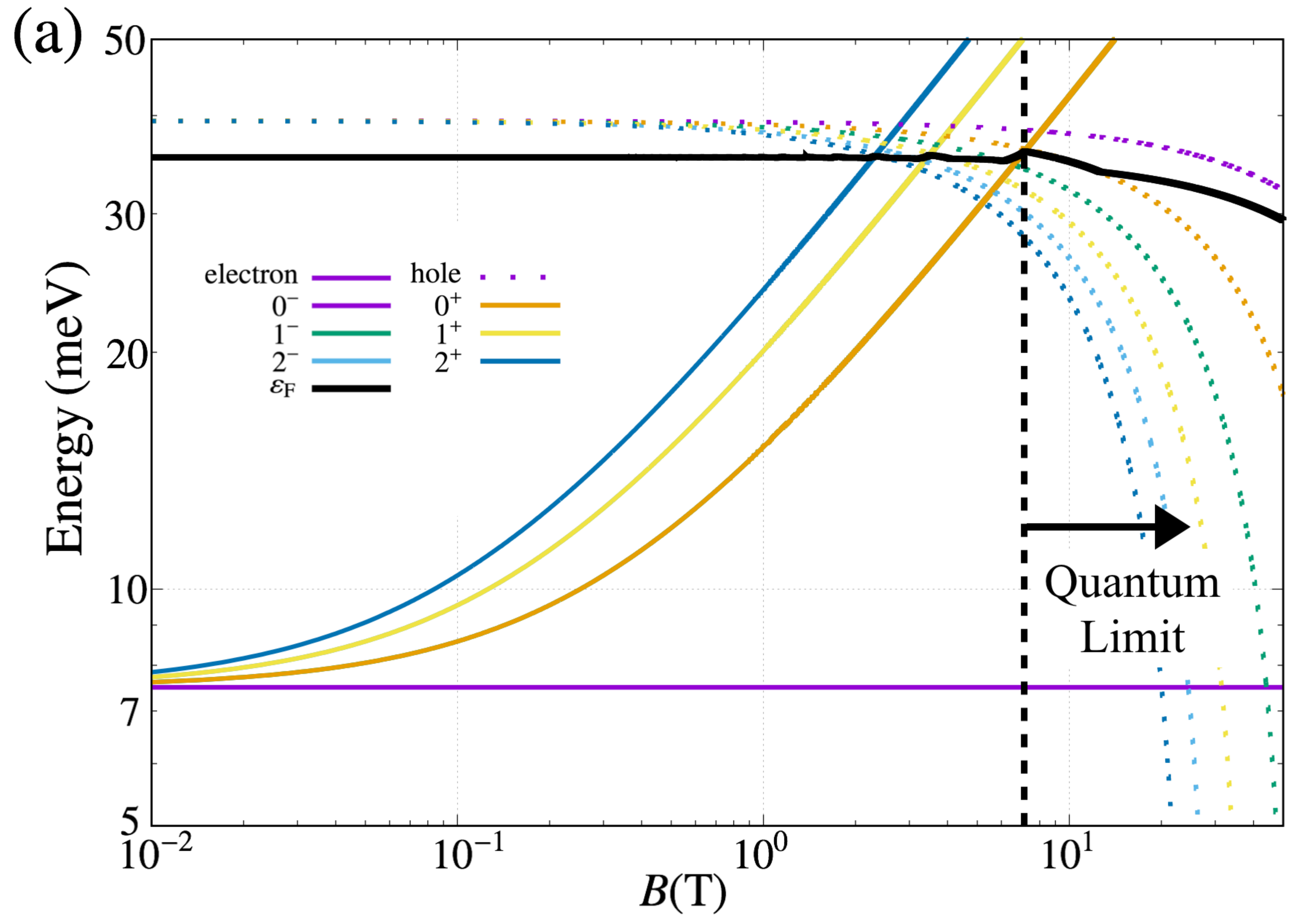}\hspace{30pt}
\includegraphics[width = 73mm]{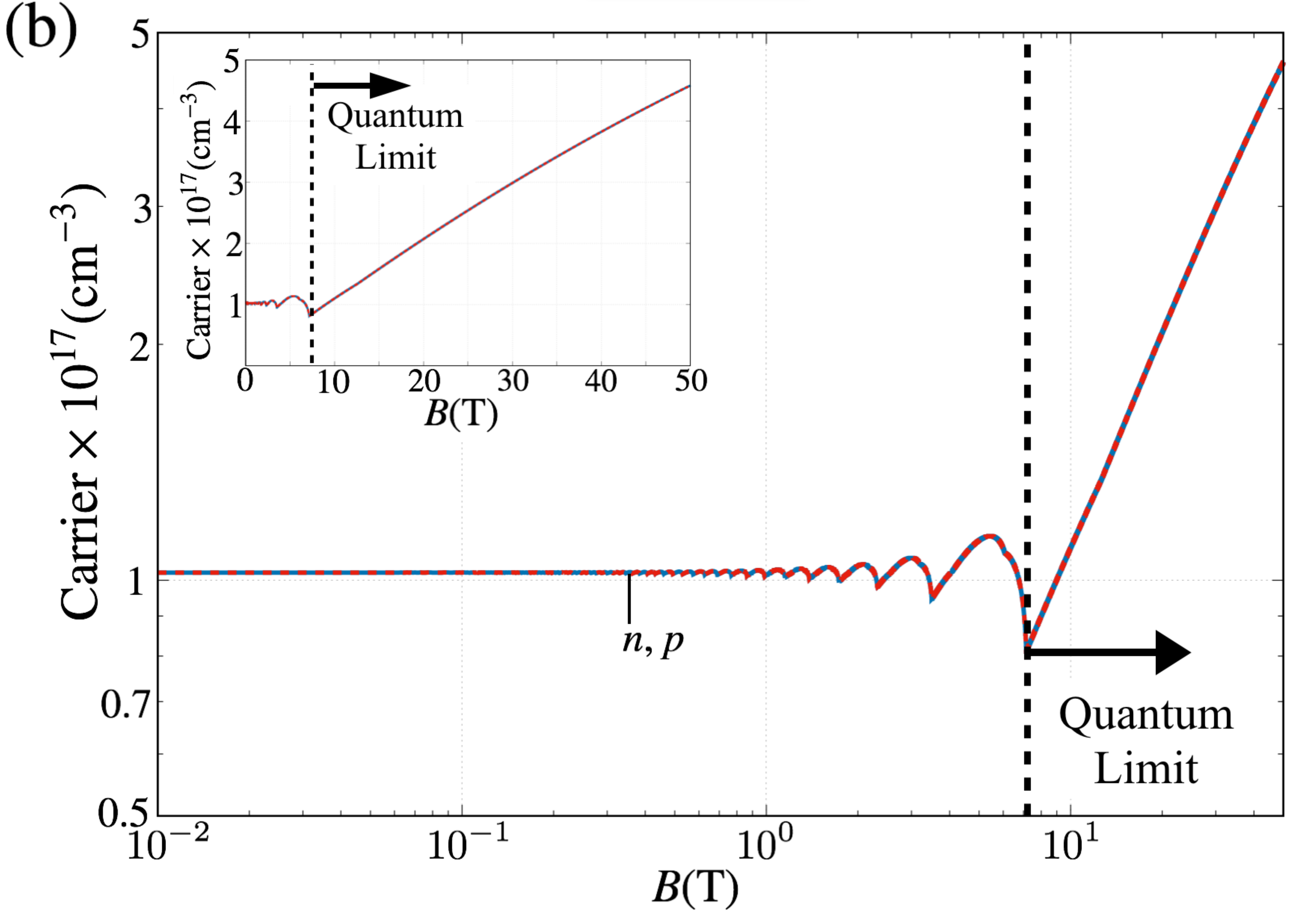}
        \caption{Field dependence of (a) the energies, and (b) the electron and hole carriers (n, p). $l^{\pm}$ on (a) is the Landau index. The solid line and broken line represent the Landau level of the electron and the hole, respectively. The solid thick line is the Fermi-energy $\ve_{\rm F}$. \label{fig_r_nr_ql}}
\end{minipage}
\end{figure*}

\begin{figure*}[t]
\begin{minipage}[]{170mm}
\centering
\includegraphics[width = 75mm]{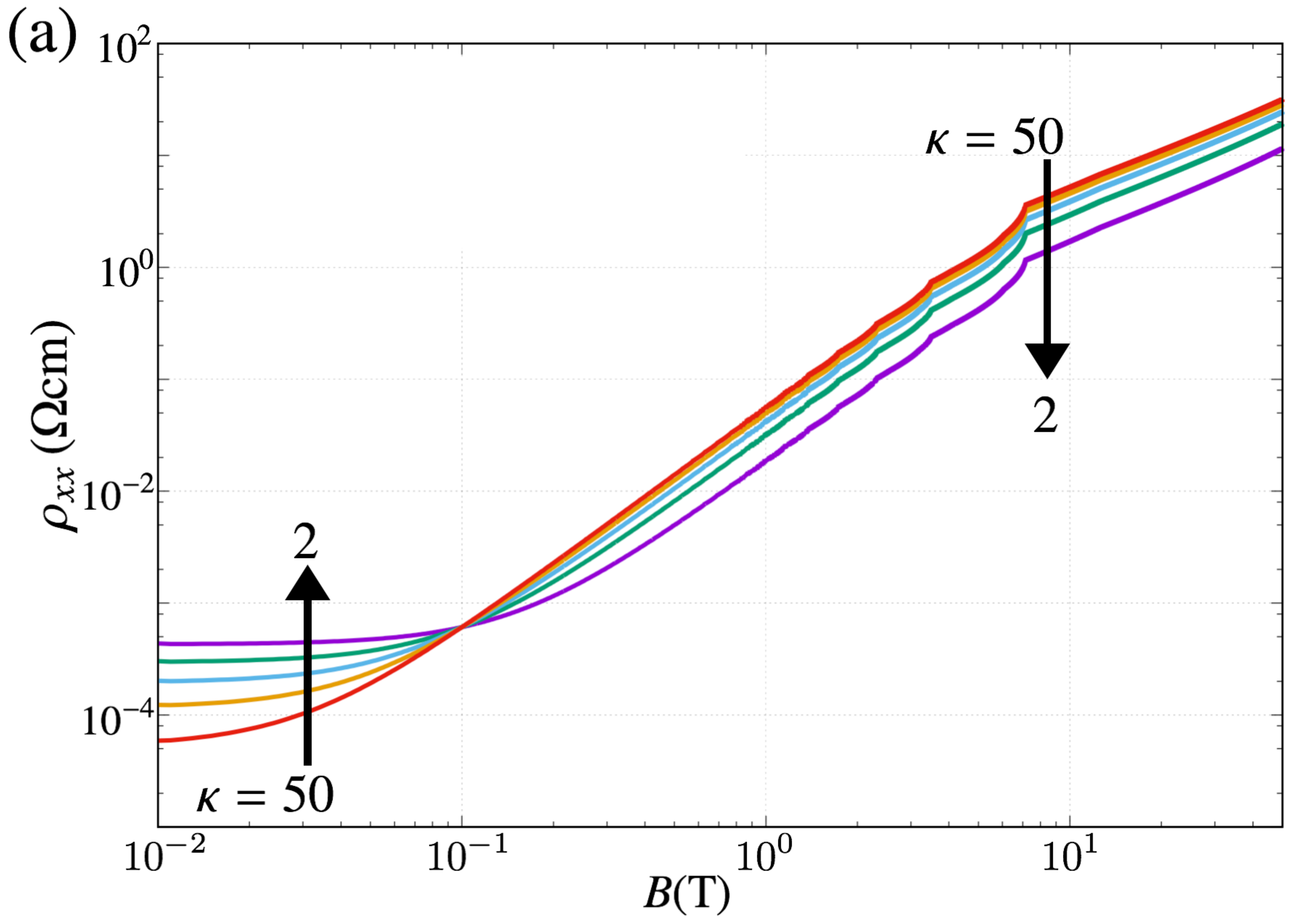} \hspace{30pt}
\includegraphics[width = 75mm]{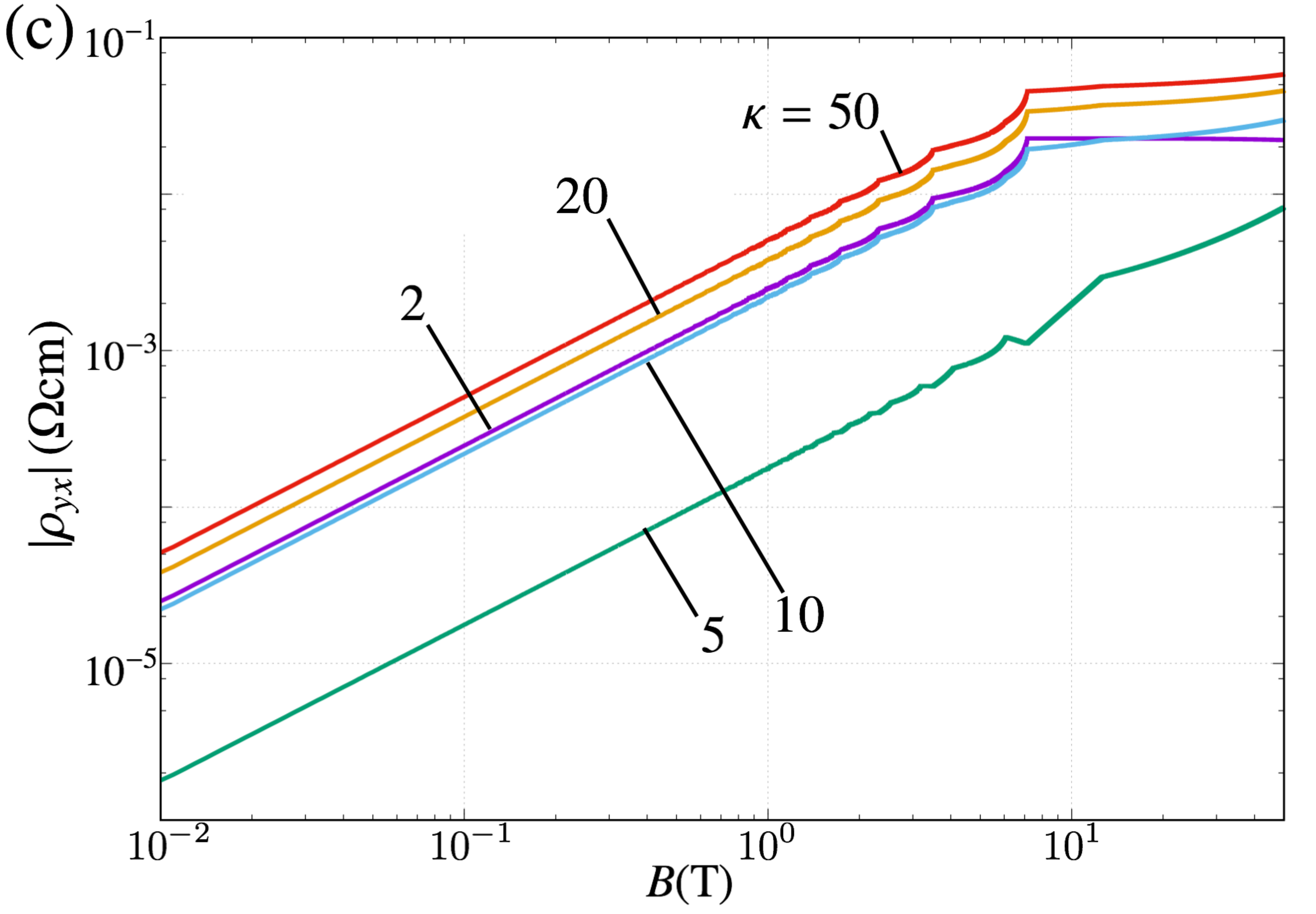}\vspace{8pt} \\
\includegraphics[width = 75mm]{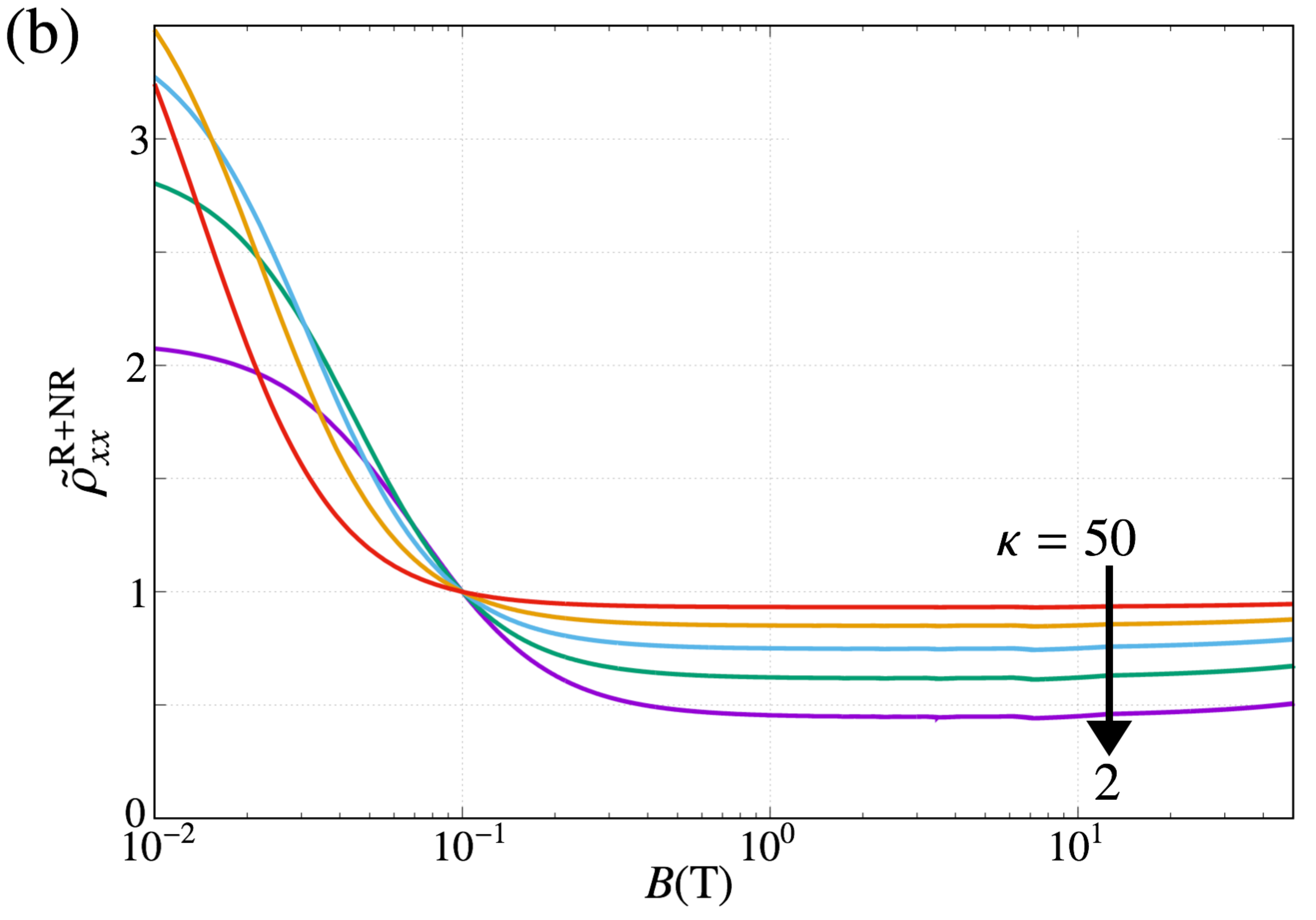}\hspace{30pt}
\includegraphics[width = 75mm]{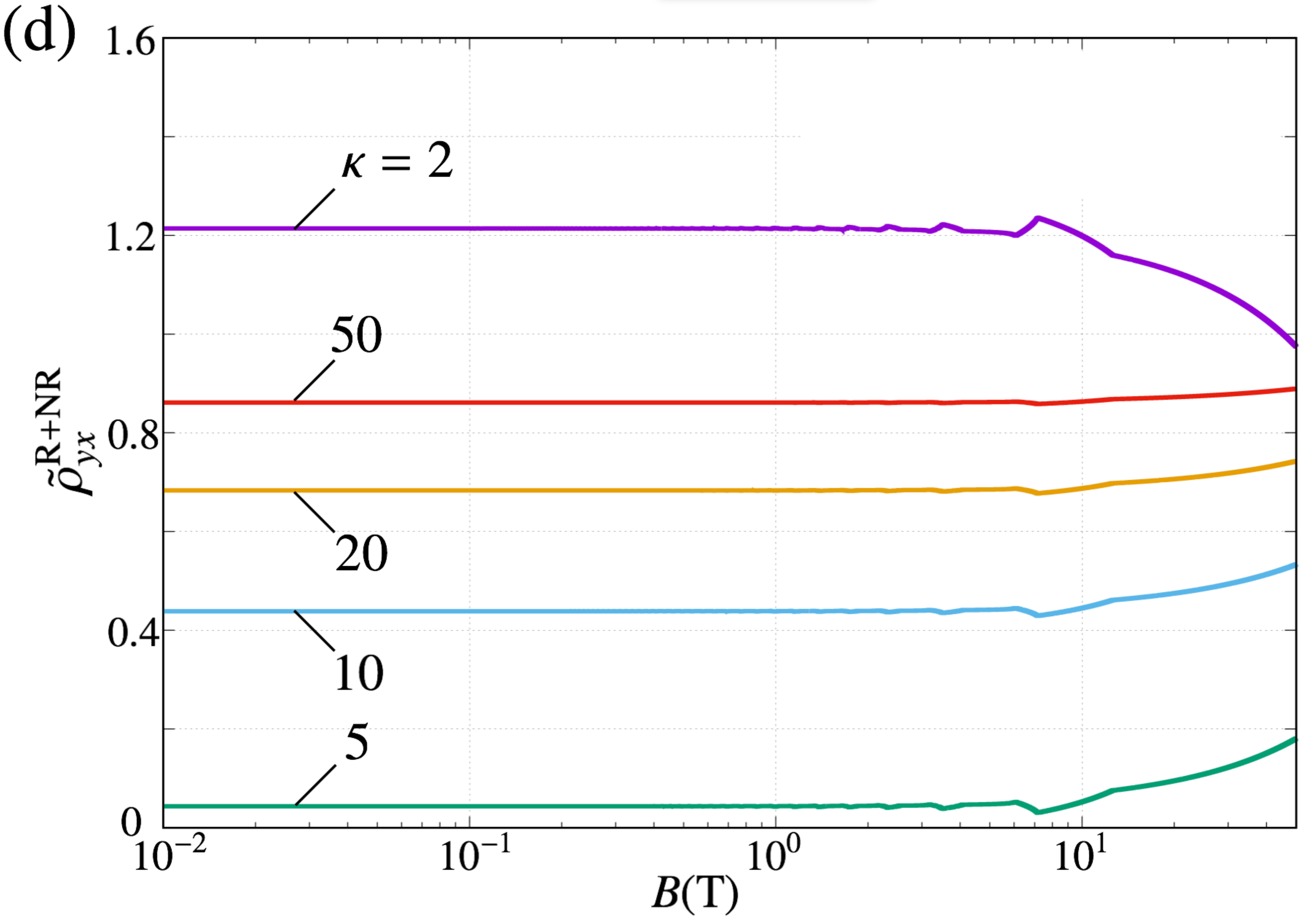}
        \caption{Field dependence of (a) the magnotoresistivity, $\rho_{xx}^{\rm R+NR}$, (b) the renormalized magnetoresistivity with the non-relativistic one, $\tilde{\rho}_{xx}^{\rm R+NR} = \rho_{xx}^{\rm R+NR}/\rho_{xx}^{\rm NR+NR}$, (c) the Hall resistivity, $\rho_{yx}^{\rm R+NR}$, and  (d) the renormalized Hall resistivity with the non-relativistic one, $\tilde{\rho}_{yx}^{\rm R+NR} = \rho_{yx}^{\rm R+NR}/\rho_{yx}^{\rm NR+NR}$ for the different ratio of $\mu_0$ to $\nu_0$, $\ka = 2, 5, 10, 20, 50$, with $\nu_0 = 10 \rm T^{-1}$ on the all figures. \label{fig_res_ql}}
\end{minipage}
\end{figure*}

The eigenvalue of holes in free electrons with the Zeeman split is \cite{SmithE1964,Zhu2011PRB}
\begin{align}
\label{eq_SBR}
\ve_0  & + \D - \ve^{\rm NR} 
=
\left(l+\frac{1}{2}\right)\hbar \Om_{\rm c} +\frac{\hbar^2 k_z^2}{2M_z} + \si \frac{G}{2}\mu_{\rm B} B,
\end{align}
where $\ve_0$, $\Om_{\rm c}$, $M_z$ and $G$ are the electron-hole hybridization, the cyclotron frequency, the longitudinal mass and the g-factor for the hole, respectively. We set $\ve_0 = 2\D$, $M_{\rm c}/m_0 = M_z/m_0 = 0.2$ and $G=5$.
$\ve_{\rm F}$ is calculated from \equsref{eq_wolff_mag} and (\ref{eq_SBR}) using the charge neutral condition $n = p$ (\figref{fig_r_nr_ql}(a)). The quantum oscillations of bismuth in the experiment can be theoretically explained with this value.
The electrons and holes reach the QL at $B \simeq 7 \rm T$ and $ B \simeq \rm 12.5T$, respectively. $\ve_{\rm F}$ in \figref{fig_r_nr_ql}(a) decreases by conserving the charge neutral condition ($n=p$). In contrast, the carrier densities $n$ and $p$ increase as the Landau degeneracy $eB/2\pi^2$ increases (\figref{fig_r_nr_ql}(b)). $\rho_{xx}^{\rm R+NR}$ and $\rho_{yx}^{\rm R+NR}$ obtained from \figref{fig_r_nr_ql} change as \figsref{fig_res_ql}(a) and (c), respectively.
As shown in \equsref{eq_res_xx2_st} and (\ref{eq_res_yx2_st}), $\rho_{xx}^{\rm R+NR}$ and $\rho_{yx}^{\rm R+NR}$ have the following dependences:
\begin{align}
\label{eq_r_linear1}
\rho_{xx}^{\rm R+NR}
 &= \frac{\la_{\ve_{\rm F}} \mu_0 \nu_0 B^2}{ne(\nu_0 + \la_{\ve_{\rm F}}\mu_0)}
\propto \frac{B^2}{n(B)}, \\
\rho_{yx}^{\rm R+NR}
 &= \frac{1}{ne}\frac{(\nu_0 - \la_{\ve_{\rm F}} \mu_0) B}{ \nu_0 + \la_{\ve_{\rm F}}\mu_0}
\propto \frac{B^1}{n(B)}. 
\end{align}
The carrier density $n$ has linear dependence $ n  \propto B^ 1$ in the QL region ($B>7 {\rm T}$), so $\rho_{xx}^{\rm R+NR}$ and $\rho_{yx}^{\rm R+NR}$ become
\begin{align}
\label{eq_r_linear2}
\rho_{xx} &\propto \frac{B^2}{n(B)} = B^1 ,\\
\label{eq_r_linear3}
\rho_{yx} &\propto \frac{B^1}{n(B)} = B^0. 
\end{align}
Now we succeeded in showing the linear MR in \equref{eq_r_linear2} only by taking into account the change of carrier density in the QL based on the semiclassical approach in contrast to the previous studies\cite{Abrikosov1969,Abrikosov2003}.
resistivity
In the last, we focus on the field dependence of the relativistic effect. \figsref{fig_res_ql}(b) and (d) show the renormalized resistivity with the non-relativistic one ($\tilde{\rho}_{ij}^{\rm R+NR} = \rho_{ij}^{\rm R+NR}/\rho_{ij}^{\rm NR+NR}$).
The relativistic effect shows three features. 
First, the relativistic effect $\la_{\ve_{\rm F}}$ increases the amplitude of magnetoresistivity ($\rho_{xx}$) under weak fields and decreases it under strong fields. 
Second, it becomes strong with a large ratio of the mobility $\ka$. 
Third,  the magnitude of the relativistic correction depends on the magnetic field in the QL, since $E_{\rm F}$ changes drastically in the QL

\section{Magnetoconductivity based on the Kubo formula}
Thus far, the MR has been discussed based on the Boltzmann theory for Dirac electrons. In this section, to check the validity of the results, we also calculate the MR based on the Kubo formula for Dirac electrons, which is represented as $\hat{\si}^{\rm KR} $.

The relativistic conductivity $\si_{\mu \nu}^{\rm KR}$ is obtained based on the Kubo formula\cite{Kubo1957_1,Nakano1956} using the relativistic correlation function $\Phi_{\mu \nu}^{\rm R}$ \cite{Fuseya2012a,Fuseya2015a} as
\begin{align}
\label{eq_kubo_si}
\si_{\mu \nu}^{\rm K} = \frac{1}{i \om} \left[ \Phi_{\mu \nu}(\om) - \Phi_{\mu \nu}(0) \right],
\end{align}
\begin{align}
\label{eq_Phi2_xx}
\Phi_{xx}^{{\rm R}} 
&= 
\frac{e^2 v^4N_{\rm L}}{8} 
\sum _{l k \si} [ f^{\rm R}_1 + f^{\rm R}_2 + f^{\rm R}_3 ] + f^{\rm R}_4,   \\
\label{eq_Phi2_yx}
\Phi_{yx}^{{\rm R} } 
&= 
\frac{e^2 v^4N_{\rm L}}{8} 
\sum _{l k \si} [ f^{\rm R}_1 - f^{\rm R}_2 - \si f^{\rm R}_3 ] + f^{\rm R}_4. 
\end{align}
Here, the Landau degeneracy $N_{\rm L} = eB/2 \pi \hbar $, $f^{\rm R}_1 , f^{\rm R}_2 , f^{\rm R}_3$ and $ f^{\rm R}_4$ are given by
\begin{align}
\label{eq_f1}
f^{\rm R}_1 
       &=       
\Xi(\om, \ve_{l\si}, \ve_{l+1\si}) \left(\Lambda_{l \si}^{l  +  1 \si}\right)^2  m_c \hbar \om_c (l + 1)      \nonumber \\
		           & \times [(\ve_{l+1\si}  +  \ve_{l\si}  +  2\D)  +  \si (\ve_{l+1\si}-\ve_{l\si})]^2 ,     \\ 
\label{eq_f2}
f^{\rm R}_2 
       &=        
\Xi(\om, \ve_{l\si}, \ve_{l-1\si}) \left(\Lambda_{l \si}^{l -1 \si}\right)^2  m_c \hbar \om_c l     \nonumber \\
		          & \times  [(\ve_{l-1\si}  +  \ve_{l\si}  +  2\D)  -  \si (\ve_{l-1\si} - \ve_{l\si})]^2 ,     \\ 
\label{eq_f3}
f^{\rm R}_3 
       &=        
2\Xi(\om, \ve_{l\si}, \ve_{l -\si})\left( \Lambda_{l \si}^{l - \si} \right)^2 \hbar^2k_z^2 (\ve_{l -\si} -  \ve_{l\si})^2 ,   \\ 
\label{eq_f4}
f^{\rm R}_4 
       &=      
  \frac{e^2 v^2 l_{\rm L}}{2} \Xi(\om)\frac{\ve_{0+} - \ve_{0-}}{\ve_{0+}}, \\
\label{eq_La}              
		&  \Lambda_a^b(\ve_a, \ve_b)   =   \frac{1}{\sqrt{\ve_a \ve_b (\ve_a  +  \D) (\ve_b  +  \D)}}.  
\end{align}
The terms $f_1$ and $f_2$ are the ``{\it orbital transition}'' and $f_3$ is the ``{\it spin transition}''. $f_4$ is the value where $\ve_a$ and $ \ve_b$ have the valence band energy of $(l,k,\si) = (0,0,-1)$. The contribution from the two one-particle Green's functions $\Xi$ becomes as follows \cite{Fuseya2015a}:
\begin{align}
\label{eq_GreenF}
        &  \Xi(\om, \ve_a, \ve_b) 
	          =\displaystyle \frac{\rm i}{2\pi} \Biggl[ \nonumber \\
	         & \displaystyle \frac{1}{\om + \ve_b - \ve_a + i \Ga} 
	           [ \ln(\ve_{\rm F}-\ve_b-i \Ga) - \ln(\ve_{\rm F} - \om -\ve_b-i \Ga)  \nonumber \\
		&          +\ln(\ve_{\rm F}-\ve_a  +i \Ga) - \ln(\ve_{\rm F} + \om -\ve_a +i \Ga) ] \nonumber \\
	        &  - \frac{1}{\om + \ve_b -\ve_a } 
	       	[ \ln(\ve_{\rm F}-\ve_b+i \Ga) - \ln(\ve_{\rm F} - \om -\ve_b-i \Ga)  \nonumber \\ 
		  &        +\ln(\ve_{\rm F}-\ve_a  -i \Ga) - \ln(\ve_{\rm F} + \om -\ve_a +i \Ga) 	] 	\Biggr],
\end{align}
where $\Ga$ is an impurity potential. The value $\Ga$ is related to $\tau$ and $\mu$ in the form
\begin{align}
\label{eq_ga}
\Ga  = \frac{\hbar}{2 \tau} = \frac{e \hbar}{2 m \mu}.
\end{align}

We set the values of $\Ga$ for electrons consistently with the calculations in the previous sections. The field dependence of energy is identical as \figref{fig_r_ql}(a).

\begin{figure}[t]
\begin{minipage}[]{87.5mm}
\centering
\includegraphics[width = 80mm]{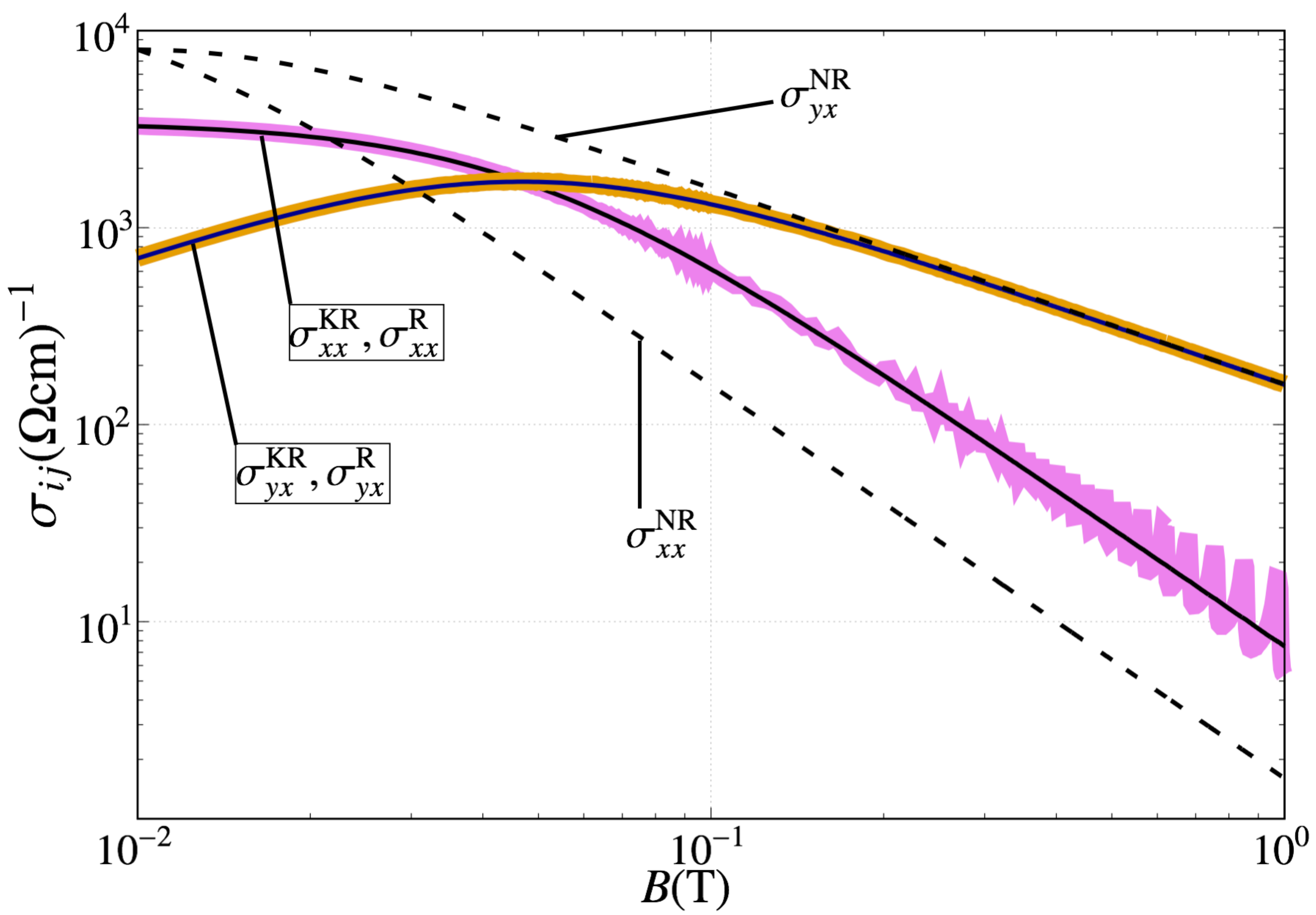}
        \caption{Field dependence of the magnetoconductivity for the relativistic and non-relativistic electron. The solid thick lines show the relativistic conductivity based on the Kubo theory, $\si_{ij}^{\rm KR}$, and the solid thin lines show it based on the Boltzmann theory, $\si_{ij}^{\rm R}$. The broken lines show the non-relativistic conductivity based on the Boltzmann theory, $\si_{ij}^{\rm NR}$. The parameters are set to be $\D = 7.5 \rm meV$, $m_{\rm c}/m_0 = 0.01$, and $\mu_0 = 100 \rm T^{-1}$. The longitudinal mass are equal to cyclotron mass ($m_{\rm c}=m_z$). \label{fig_cdv_bvsk_r_weak}}
\end{minipage}
\end{figure}

Magnetoconductivity so obtained, $\si_{{\rm }ij}^{\rm KR}$, is shown in \figref{fig_cdv_bvsk_r_weak} for $\D = 7.5\rm meV$, $m_c/m_0 = m_z/m_0 = 0.01$, $\mu_0 = 100{\rm T^{-1}}$. 
As is seen in \figref{fig_cdv_bvsk_r_weak}, there is a clear gap between the results by the Kubo formula $\si_{ij}^{\rm KR}$ (solid thick line) and those by Boltzmann theory for non-relativistic electron $\si_{ij}^{\rm NR}$ (broken line). On the other hand, $\si_{ij}^{\rm R}$ (solid thin line) agrees with $\si_{ij}^{\rm KR}$ other than the quantum oscillation in $\si_{ij}^{\rm KR}$. This implies that our formula of $\si_{ij}^{\rm R}$ with the relativistic correction $\la_{\ve_{\rm F}} ( = \D/\ve_{\rm F})$ gives correct results under any magnitude the magnetic field even though it is the semiclassical approach. 

The field dependence of $\hat{\si}^{\rm KR}$, the conductivity based on the Kubo theory, is vague, and so is $\hat{\rho}^{\rm KR}$. While, the field dependence of our $\hat{\si}^{\rm R}$ is very clear. $\hat{\si}^{\rm R}$ is more useful instrument for analysis of experimental data.

\section{Summary}
We studied the carrier transport in magnetic fields for relativistic electrons. 
We found the new factor $\la_{\ve_{\rm F}} = \D/\ve_{\rm F}$, the relativistic correction, that is not appear in the conventional non-relativistic formula. The effect of the relativistic correction is different between at weak fields and at strong fields. Furthermore, $\la_{\ve_{\rm F}}$ depends on the magnetic field near the QL. The MR of the Dirac electron system should be analyzed by the relativistic resistivity formula ($\hat{\rho}^{\rm R}, \hat{\rho}^{\rm R+NR}$). In our formula, the field dependence of magneto-`resistivity' is clearly indicated. This makes the analysis of experiments more transparent.

The gap between the semi-classical and quantum approaches for the conductivities of relativistic electrons is removed. The validity of our formula is verified by the calculation based on the Kubo formula for relativistic electrons. Therefore, the magnetoresistance of Dirac electrons can be quantitatively evaluated only with the semiclassical formula. 

As a byproduct, the linear MR of bismuth is qualitatively explained by taking into account dependence of the carrier density.

In the previous studies\cite{Collaudin2015,Emoto2016,Zhu2017}, the MR of bismuth has been analyzed by using the conventional formula of the non-relativistic carriers for two bands. The analysis can be quantitatively corrected by considering the relativistic correction found in the present work. In order to obtain quantitative evaluation of MR of bismuth, we further need to take into account the anisotropy of the mobility, which will be published elsewhere near future.

\section*{Acknowledgement}
We thank K. Behnia, B. Fauqu\'e and Z. Zhu for fruitful discussions. This work was supported by JSPS KAKENHI grants 16K05437, 15KK0155 and 15H02108.

\bibliographystyle{jpsj}
\bibliography{ref_jpsj}

\end{document}